\documentclass[traditabstract]{aa}
\usepackage{epsfig}
\usepackage{aalongtable}
\usepackage{natbib,twoopt}
 \bibpunct{(}{)}{;}{a}{}{,}    %% natbib format like A&A and ApJ
\usepackage{graphics}
\usepackage{color} 
\usepackage{ulem}
\usepackage{times}
\usepackage{color}
\newfont{\tss}{cmssdc10 scaled 950}

\begin{document}

\title{Dust emission in star-forming dwarf galaxies: General properties and
the nature of the sub-mm excess
\thanks{Figure \ref{fig6} is available only in the electronic edition.}
}
% and at the CDS via
%anonymous ftp to cdsarc.u-strasbg.fr (130.79.128.5) or via
%http://cdsweb.u-strasbg.fr/cgi-bin/qcat?J/A+A/?/?}

%\subtitle{}

\author{Y. I.\ Izotov \inst{1,2,3}
\and N. G.\ Guseva \inst{1,2}
\and K. J.\ Fricke \inst{1,4}
%\and G.\ Stasi\'nska \inst{3}
\and E.\ Kr\"ugel \inst{1}
\and C.\ Henkel \inst{1,5}
%\and P.\ Papaderos \inst{5,6}
}
\offprints{Y.I. Izotov, izotov@mao.kiev.ua}
\institute{          Max-Planck-Institut f\"ur Radioastronomie, 
                     Auf dem H\"ugel 
                     69, 53121 Bonn, Germany
\and
                     Main Astronomical Observatory,
                     Ukrainian National Academy of Sciences,
                     Zabolotnoho 27, Kyiv 03680,  Ukraine
\and
                     LUTH, Observatoire de Paris, CNRS, 
                     Universite Paris Diderot,
                     Place Jules Janssen 92190 Meudon, France
\and
                     Institut f\"ur Astrophysik, 
                     G\"ottingen Universit\"at, Friedrich-Hund-Platz 1, 
                     37077 G\"ottingen, Germany
\and
                     Astronomy Department, King Abdulaziz University, 
                     P.O. Box 80203, Jeddah 21589, Saudi Arabia
%\and
%                     Department of Astronomy, Oskar Klein Centre, Stockholm 
%                     University, SE - 106 91 Stockholm, Sweden
}

\date{Received \hskip 2cm; Accepted}
\abstract{We studied the global characteristics of dust emission in a large 
sample of emission-line star-forming galaxies. The sample consists
of two subsamples. One subsample (SDSS sample) includes $\sim$4000 
compact star-forming galaxies from the Sloan Digital Sky Survey (SDSS),
which were also detected in all four bands at 3.4$\mu$m, 4.6$\mu$m,
12$\mu$m, and 22$\mu$m of the WISE all-sky survey. The second subsample
(Herschel sample) is a sample of 28 compact 
star-forming galaxies observed with Herschel in the FIR range. 
Data of the Herschel sample were supplemented by the photometric data 
from the Spitzer observations, Galaxy Evolution Explorer (GALEX) 
survey, Sloan Digital Sky Survey (SDSS), 
Wide-Field Infrared Survey Explorer (WISE), Two Micron All Sky Survey 
(2MASS), NRAO VLA Sky Survey 
(NVSS), and Faint Images of the Radio Sky at Twenty-cm (FIRST) survey, as 
well as optical and 
Spitzer spectra and data in sub-mm and radio ranges.
It is found that warm dust luminosities 
of galaxies from the SDSS sample and cold and warm dust luminosities
of galaxies from the Herschel sample are strongly correlated with 
H$\beta$ luminosities, which implies that one of the main sources of dust 
heating in star-forming galaxies is ionising UV radiation of young stars. 
Likely, a significant fraction of dust is located inside H~{\sc ii} and
surrounding regions. We found tight correlations between masses
of cold and warm dust, again implying the same source of dust heating.
%, UV radiation of star-forming regions. 
Using the relation between warm and cold dust masses for 
estimating the total dust mass in star-forming galaxies with an accuracy
better than $\sim$ 0.5 dex is proposed.
On the other hand, it is shown for both samples that 
dust temperatures do not depend on the metallicities.
% is not a factor, which regulates dust emission.
The dust-to-neutral gas mass ratio strongly declines with decreasing
metallicity, similar to that found in other studies of local
emission-line galaxies, high-redshift gamma-ray burst (GRB) hosts, and 
damped Ly-$\alpha$ absorbers (DLAs). On the other hand, the 
dust-to-ionised gas mass ratio is about one hundred times as high
implying that most of dust is located in the neutral gas.
It is found that thermal free-free emission of ionised gas in
compact star-forming galaxies is important in the sub-mm and mm
ranges, and it might be responsible for the sub-mm emission excess. 
This effect is stronger in galaxies with lower
metallicities and is also positively affected by an increased star-formation
rate.
}
\titlerunning{Dust emission in star-forming dwarf galaxies}
\authorrunning{Y.I.Izotov et al.}
\maketitle

%\markboth{Y.I.Izotov et al.}{}

\section{Introduction}

Dust plays an important role in the thermal and dynamical evolution of galaxies
at different spatial scales from protostars to large molecular complexes.
Observations with ground-based and space-born telescopes, Infrared Astronomical 
Satellite (IRAS), Infrared Space Observatory (ISO),
Submillimetre Common-User Bolometer Array ({\it SCUBA}), Spitzer, 
and Herschel, revealed significant
populations of dusty luminous and ultraluminous infrared galaxies at redshifts
$z$$\ga$1 \citep[e.~g. ][]{HD01,E02,D06,F06}, indicating that 
dust grains are being formed in intense and short star-formation episodes 
and implying a rapid transition in high-redshift galaxies
from the formation of zero-metal Population III stars to the formation of 
stellar populations more typical for present-day galaxies 
\citep[e.g. ][]{N03,S12,C14}. However, the dust formation in the early
universe is debated. In particular, \citet{V11} find that population III stars
do not contribute much to the dust budget. They claim that the dust is formed
by population II/I, SNe, and AGB stars at the later stage.

Studies of high-redshift low-metallicity galaxies are very difficult 
because of their faintness and small angular sizes. We have a chance to
study star formation and role of dust in low-metallicity environments 
by investigating properties of nearby star-forming dwarf galaxies, such as
irregular, blue compact dwarf (BCD), and green pea (GP) galaxies. 
These galaxies can be considered as local counterparts or ``analogs'' of
high-redshift galaxies, such as Lyman-break galaxies (LBGs) and Lyman-$\alpha$
emitting galaxies, because of their high star-formation activity and
low metallicity \citep[e.g.][]{C09,I11a}. 
Due to their proximity they can be
studied in greater detail than it is possible in distant galaxies.
Studying the properties of a metal-poor ISM, 
we may be able to better understand the physical conditions in high-redshift
galaxies.

The properties of radiation in the mid- and 
far-infrared ranges of star-forming dwarf galaxies 
were studied in many papers by considering samples of
galaxies and individual galaxies
\citep[e.g. ][]{D07,E05,E08,C06,H05,H10,G08,Ga09,Ga11,M06,M13,RR13,RR14,W06,W07,W08,H09,O10,C12,In09,In13,K13}. Spectral energy distributions (SEDs) 
in these ranges are dominated by dust emission resulting from reprocessing of
UV radiation from young massive stars, in particular in the 
most metal-deficient BCDs I~Zw~18 and SBS~0335$-$052E \citep{W07,H04}.

It was concluded in numerous papers that the main dust heating source in 
dwarf emission-line galaxies is the radiation from young
stellar populations. Dust in these galaxies is heated to higher temperatures
of $\sim$ 30-40K as compared to $\sim$ 20K in spirals 
\citep[e.g. ][]{E08,H08}. The polycyclic aromatic hydrocarbon (PAH) emission 
in dwarf star-forming galaxies is weak or absent, and it decreases with 
decreasing metallicity \citep[e. g. ][]{W07,H10}. Different mechanisms were
invoked to explain the weakness of PAH features in these galaxies:
destruction by the intense UV radiation \citep{M06}, 
destruction by shocks \citep{O06}, delayed formation by AGB stars \citep{G08},
and formation of molecular clouds \citep{Sa10}.
It was also found that the dust emission in dwarf galaxies cannot
be characterised by radiation with a single temperature. 
Warm and hot dust with temperatures of up to several hundred degrees must be 
present in many of these galaxies in addition to cold dust to 
reproduce the SED in the mid- and far-infrared ranges 
\citep[e. g. ][]{G11,I11b,I14}.
It is most likely that dust temperature gradients are present in the ISM 
around massive and compact stellar clusters \citep[e.g. ][]{H13}.

The mid- and far-infrared emission in dwarf galaxies results from
absorption of radiation at shorter wavelengths in the UV and optical
ranges. It includes not only the stellar component and nebular
emission lines but also the nebular continuum (free-bound and
free-free emission). The necessity to
include the latter in young starbursts was discussed by
\citet{I11a} as otherwise stellar masses may be overestimated by a
factor of up to $\sim$ 3.

It also was established that the fraction of nebular continuum
increases with wavelength.  Nebular continuum dominates at
$\lambda$$\lambda$ 1 - 3 $\mu$m with a flux fraction of 70-90\% in
extremely young starbursts, which are characterised by high equivalent
widths EW(H$\beta$) $>$100\AA\ of the H$\beta$ emission line.
Ignoring this result would lead to wrong conclusions. For example,
\citet{E08} suggested that dust already dominates at 1.8$\mu$m in 
SBS~0335-052E, Haro~11, and SHOC~391, while it is clear from detailed
fitting of optical spectra that the main contributor to the continuum
in the near-infrared range is ionised gas.

Therefore, photometric data are not sufficient for correct SED fitting
in the optical range; spectra are needed as well.  Using spectra and the
H$\beta$ emission line flux, one can also predict the flux of free-free
emission in the radio range \citep{CD86} and can estimate how much star
formation is hidden in the UV and optical ranges.

Among the first studies, where detailed fitting of UV and optical SEDs, which
included both stellar and gaseous continua, was applied for analysis of
optical and infrared properties of large samples of compact star-forming 
galaxies were those performed by \citet{I11b} and \citet{I14}. 
In particular, \citet{I14} considered a sample
of $\sim$ 14000 SDSS galaxies with strong emission lines and found that
mid- and far-infrared luminosities are strongly correlated with the
luminosity of ionising stellar radiation.

This study is the continuation of the work by \citet{I11b} and \citet{I14},
where relations between luminosities and optical and near-infrared
colour characteristics were mainly discussed. Here we concentrate on the 
determination of the dust characteristics, such as temperature and
mass. We use two samples. The first sample selected from the SDSS 
consists of compact star-forming galaxies
with available mid-infrared data from the WISE all-sky survey.
This sample was discussed by \citet{I14}. 
The second sample consists of nearby
star-forming galaxies observed with Herschel by \citet{M13} 
and \citet{RR13}.

The criteria for selection and properties of both samples are discussed
in Sect. \ref{samples}. We describe the determination of galaxy parameters
in Sect. \ref{param}. Results of this study are discussed in Sect. \ref{results}.
Our findings are summarised in Sect. \ref{sum}.

\section{The samples \label{samples}}

%To analyse the integrated properties of dust emission in star-forming
%galaxies, we use two samples. The first sample is the sample selected from the 
%SDSS (SDSS sample) \citep{I14} and the second sample is 
%the sample of star-forming galaxies studied by \citet{RR13} with 
%the Herschel space observatory (Herschel sample). 

\subsection{SDSS sample}

   We use a sample of $\sim$ 14000 star-forming galaxies selected by 
\citet{I14} from the spectroscopic database of the SDSS Data Release 7
(DR7) \citep{A09}. The details of data selection can be found in \citet{I14}.
Out of 
%the \citet{I14} 
this sample,  we select $\sim$ 4000 compact
galaxies with angular diameters $\leq$6\arcsec, 
which were detected in all four WISE bands at 
$\lambda$3.4$\mu$m, $\lambda$4.6$\mu$m, $\lambda$12$\mu$m, and 
$\lambda$22$\mu$m, allowing the determination of the warm and hot
dust properties. 
The SDSS spectra and SDSS $u$, $g$, $r$, $i$, and $z$ magnitudes of all 
selected galaxies are available in the SDSS database.
Finally, we restrict the sample to only those $\sim$ 1000 galaxies out 
of the $\sim$ 4000, where the [O {\sc iii}] $\lambda$4363\AA\ emission line 
was measured with an accuracy
better than 50\%. This allows for a reliable determination of the oxygen 
abundance. Hydrogen emission lines in these galaxies are strong allowing for a
correction for dust extinction from the decrement of several hydrogen lines.
%Thus, the star formation activity in $\sim$ 1000 selected galaxies is high.

%\textbf{The selected SDSS objects were cross-identified with
%sources from GALEX (UV), 2MASS (NIR), WISE (MIR),
%IRAS (FIR) and NVSS (radio) photometric sky surveys, fully covering the 
%range of equatorial coordinates of the SDSS.} ****** remove???

\subsection{Herschel sample}\label{Herschel}

We use a sample of 48 star-forming galaxies studied by \citet{RR13} with the
Herschel space observatory and adopt their Herschel
FIR fluxes and distances. This sample was supplemented by the data 
on the total fluxes from other photometric surveys, namely from the
GALEX\footnote{http://galex.stsci.edu/GR6/} \citep{M05}, 
SDSS\footnote{http://skyserver.sdss3.org/dr10/en/tools/chart/chartinfo.aspx}
\citep{A09}, WISE\footnote{http://irsa.ipac.caltech.edu/cgi-bin/Gator/nph-scan?mission=irsa\&submit=Select\&projshort=WISE} \citep{W10}, 2MASS\footnote{http://irsa.ipac.caltech.edu/cgi-bin/Gator/nph-scan?mission=irsa\&submit=Select\&projshort=2MASS} \citep{S06}, 
NVSS\footnote{http://www.cv.nrao.edu/nvss/} \citep{C98} and
FIRST\footnote{http://sundog.stsci.edu/cgi-bin/searchfirst} \citep{KI08}
surveys. 
Furthermore, Spitzer spectra\footnote{http://cassis.sirtf.com/atlas/index.shtml} \citep{L11} in the 
mid-infrared range, spectra in the optical range, and, whenever it was 
possible, data in the sub-mm and radio ranges at different wavelengths 
were collected. Out of 48 galaxies by \citet{RR13}, we
excluded galaxies without optical or Spitzer spectra, galaxies
with large angular radii $\ga$60\arcsec, and galaxies, which were not detected
by Herschel in any of the {\it PACS} or {\it SPIRE} photometric bands. 
Furthermore, we
excluded the compact object UM 311 because this is not a galaxy, but 
an H~{\sc ii} region within a spiral galaxy.

In the end, the Herschel sample consists of 28 compact galaxies with 
coordinates and distances shown in Table \ref{tab0}. References on 
available photometric and spectroscopic data are also given in the Table.

\renewcommand{\baselinestretch}{1.0}

 \begin{table*}
 \caption{References to the data for galaxies from the {\sl Herschel} sample.}
 \label{tab0}
 \begin{tabular}{lcccccccc} \hline \hline
 & & & &\multicolumn{5}{c}{References} \\ \cline{5-9}
Object         &RA (J2000) &Dec (J2000)   &Distance&H$\beta$&Optical&Spitzer   &sub-mm and&H {\sc i} 21cm \\
               &           &              &        &flux    &spectra&photometry&radio     &               \\ \hline
Haro 11        &00h36m52.7s&$-$33d33m17.0s& 92.1   &  1     &  2    & 3        & 4,1,5,6          & 53 \\
Haro 3         &10h45m22.4s&$+$55d57m37.0s& 19.3   &  7     &  8    & 9        & 5,54,57          & 10  \\
HS 0017$+$1055 &00h20m21.4s&$+$11d12m21.0s& 79.1   & 11     & 11    & ...      & 57               & ... \\
HS 0052$+$2536 &00h54m56.4s&$+$25d53m08.0s& 191.0  & 11     & 11    & ...      & ...              & ... \\
HS 0822$+$3542 &08h25m55.5s&$+$35d32m32.0s& 11.0   &  7     & 12    & 3        & 57               & 13  \\
HS 1222$+$3741 &12h24m36.7s&$+$37d24m37.0s& 181.7  & 12     & 12    & ...      & 57               & ... \\
HS 1304$+$3529 &13h06m24.2s&$+$35d13m43.0s& 78.7   & 12     & 12    & ...      & 57               & ... \\
HS 1330$+$3651 &13h33m08.3s&$+$36d36m33.0s& 79.7   & 12     & 12    & ...      & 57               & ... \\
I Zw 18        &09h34m02.0s&$+$55d14m28.0s& 18.2   &  7     & 12    & 3        & 14,16,57         & 10  \\
II Zw 40       &05h55m42.6s&$+$03d23m32.0s& 12.1   & 17     & 18    & 3        & 19,20,21,22,23,5,54 & 24  \\
Mrk 1089       &05h01m37.7s&$-$04d15m28.0s& 56.6   & 25     & 26    & ...      & 1,5,27,54        & 29  \\
Mrk 1450       &11h38m35.7s&$+$57d52m27.0s& 19.8   &  7     & 30    & 3        & 5,57             & 31  \\
Mrk 153        &10h49m05.0s&$+$52d20m08.0s& 40.3   &  7     & 12    & 3        & 5,57             & 32  \\
Mrk 209        &12h26m15.9s&$+$48d29m37.0s&  5.8   &  7     & 12    & 33       & 5,55,56,57       & 10  \\
Mrk 930        &23h31m58.3s&$+$28h56m50.0s& 77.8   &  7     &  8    & 3        & 34,5             & 35  \\
NGC 1140       &02h54m33.6s&$-$10d01m40.0s& 20.0   &  7     & 18    & 3        & 19,20,28,5       & 24  \\
Pox 186        &13h25m48.6s&$-$11d36m38.0s& 18.3   & 37     & 37    & ...      & ...              & ... \\
SBS 0335$-$052E&03h37m44.0s&$-$05d02m40.0s& 56.0   & 38     & 39    & 3        & 40,41,5,54       & 42  \\
SBS 1159$+$545 &12h02m02.4s&$+$54d15m50.0s& 57.0   & 30     & 30    & ...      & 57               & ... \\
SBS 1211$+$540 &12h14m02.5s&$+$53d45m17.0s& 19.3   & 30     & 12    & ...      & 57               & 35  \\
SBS 1249$+$493 &12h51m52.5s&$+$49d03m28.0s& 110.8  & 43     & 43    & ...      & 57               & 52  \\
SBS 1415$+$437 &14h17m01.4s&$+$43d30m05.0s& 13.6   & 44     & 43    & ...      & 57               & 35  \\
SBS 1533$+$574 &15h34m13.8s&$+$57d17m06.0s& 54.2   & 45     & 45    & ...      & 57               & 35  \\
Tol 1214$-$277 &12h17m17.1s&$-$28d02m33.0s& 120.5  & 46     & 47    & 3        & ...              & ... \\
UGC 4483       &08h37m03.0s&$+$69d46m31.0s&  3.2   &  7     & 30    & 3        & ...              & 48  \\
UM 448         &11h42m12.4s&$+$00d20m03.0s& 87.8   &  7     & 39    & 3        & 28,49,5,57,58    & 10  \\
UM 461         &11h51m33.3s&$-$02d22m22.0s& 13.2   &  7     & 12    & 3        & 57               & 50  \\
VII Zw 403     &11h27m59.9s&$+$78d59m39.0s&  4.5   &  7     & 45    & 3        & 51,59            & 10  \\
 \hline
 \end{tabular}
   
\noindent {\sl Note}: Distances and {\sl Herschel} photometric data are from \citet{RR13}. 
{\sl GALEX} photometric data are from
http://galex.stsci.edu/GR6/. SDSS photometric data are from 
http://skyserver.sdss3.org/dr10/en/tools/chart/chartinfo.aspx. {\sl WISE} data are from
http://irsa.ipac.caltech.edu/cgi-bin/Gator/nph-scan?mission=irsa\&submit=Select\&projshort=WISE.
2MASS data are from http://irsa.ipac.caltech.edu/cgi-bin/Gator/nph-scan?mission=irsa\&submit=Select\&projshort=2MASS.
NVSS and FIRST data in the continuum at 20 cm are from http://www.cv.nrao.edu/nvss/ and http://sundog.stsci.edu/cgi-bin/searchfirst,
respectively; low-resolution {\sl Spitzer} spectra are from http://cassis.sirtf.com/atlas/index.shtml.

\noindent {\sl References}: (1) \citet{Sc06}; (2) \citet{G12}; (3) \citet{E08}; (4) \citet{Ga09}; (5) \citet{C98};
(6) \citet{M06}; (7) \citet{MK06}; (8) \citet{IT04}; (9) \citet{H06}; (10) \citet{HR89}; (11) \citet{U03};
(12) \citet{I06}, http://das.sdss.org/spectro/; (13) \citet{Ch06}; (14) \citet{Le07}; (15) \citet{C05}; (16) \citet{H05b};
(17) \citet{La07}; (18) \citet{G00}; (19) \citet{G05}; (20) \citet{A04}; (21) \citet{H11}; (22) \citet{H05b};
(23) \citet{K14}; (24) \citet{Sp05}; (25) \citet{VC92}; (26) \citet{IT99}; (27) \citet{Ga09}; (28) \citet{B95};
(29) \citet{W91}; (30) \citet{ITL94}; (31) \citet{Hu05}; (32) \citet{TM81}; (33) \citet{D09}; (34) \citet{R07};
(35) \citet{T99b}; (36) \citet{H94}; (37) \citet{G04}; (38) \citet{I06b}; (39) \citet{IT98}; (40) \citet{J09};
(41) \citet{H13}; (42) \citet{E09}; (43) \citet{TIL95}; (44) \citet{GP03}; (45) \citet{ITL97}; (46) \citet{Gu11};
(47) \citet{I01}; (48) \citet{dV91}; (49) \citet{DC78}; (50) \citet{Co11}; (51) \citet{THL04}; (52) \citet{P02};
(53) \citet{Co14}; (54) \citet{R14}; (55) \citet{V83}; (56) \citet{K91}; (57) \citet{KI08}; (58) \citet{J02};
(59) \citet{Le05}.

 \end{table*}

\section{The determination of galaxy parameters \label{param}}

Most global galaxy parameters (metallicities, stellar masses, H$\beta$
luminosities and UV, optical, infrared and radio monochromatic luminosities)
for the SDSS sample were derived by
\citet{I14}, which we adopt in this paper. We apply the same technique to
derive galaxy parameters for the Herschel sample. The technique is
briefly described in this section.

\subsection{Element abundances and H$\beta$ luminosity \label{abund}}

The SDSS spectra of the galaxies from the SDSS sample are used
%were used 
to derive emission-line fluxes and equivalent widths, the extinction
coefficient $C$(H$\beta$), the H~{\sc ii} region luminosity in the H$\beta$ 
emission line, and chemical element abundances. The observed line
fluxes were obtained using the IRAF\footnote {IRAF is the Image 
Reduction and Analysis Facility distributed by the National Optical Astronomy 
Observatory, which is operated by the Association of Universities for Research 
in Astronomy (AURA) under cooperative agreement with the National Science 
Foundation (NSF).} SPLOT routine. The line flux errors 
included statistical errors in addition to errors introduced by
the standard star absolute flux calibration, which we set to 1\% of the
line fluxes. These errors are later propagated into the calculation
of abundance errors.
The line fluxes were corrected for the Milky Way and internal reddening using 
the extinction curve of \citet{C89} and for underlying hydrogen stellar 
absorption. The extinction coefficients are defined as 
$C$(H$\beta$) = 1.47$E(B-V)$, where $E(B-V)$ = $A(V)$/$R_V$ and $R_V$ = 
3.2 \citep{A84}. 

   The oxygen abundances were derived according to the procedures 
described by \citet{ITL94,ITL97} and \citet{TIL95}, where a two-zone 
photoionised H~{\sc ii} region model was adopted: a high-ionisation zone with 
temperature $T_{\rm e}$(O~{\sc iii}),
and a low-ionisation zone with temperature $T_{\rm e}$(O~{\sc ii}).
In the H~{\sc ii} regions with a detected [O~{\sc iii}] $\lambda$4363
emission line, the temperature $T_{\rm e}$(O~{\sc iii}) was calculated using 
the direct method based on the 
[O~{\sc iii}] $\lambda$4363/($\lambda$4959+$\lambda$5007) line ratio.
The electron temperature $T_{\rm e}$(O~{\sc iii})
in H~{\sc ii} regions with non-detected 
[O~{\sc iii}] $\lambda$4363 emission line was derived by a semi-empirical
method \citep{IT07,I14}, which is based on the relation between 
$T_{\rm e}$(O~{\sc iii}) and the total flux of strong 
[O {\sc ii}] $\lambda$3727, [O {\sc iii}] $\lambda$4959, and
[O {\sc iii}] $\lambda$5007 emission lines.

We use the relation between the electron temperatures 
$T_{\rm e}$(O~{\sc iii}) and $T_{\rm e}$(O~{\sc ii}) obtained by \citet{I06} 
from
the H~{\sc ii} region models of \citet{SI03} to derive 
$T_{\rm e}$(O~{\sc ii}). Ionic and total oxygen abundances were derived
using expressions for ionic abundances and ionisation correction factors 
(ICFs) obtained by \citet{I06}.

The extinction-corrected luminosity $L$(H$\beta$) was obtained from the
observed H$\beta$ emission-line flux by adopting the total extinction, which
included the Milky Way, internal galaxy extinctions, and the
distance derived from the redshift. For distance determination, 
\citet{I14} used the relation 
$D$ = $f$($z$,$H_0$,$\Omega_{\rm M}$,$\Omega_\Lambda$)
from \citet{R67}, where
the Hubble constant $H_0$ = 67.3 km s$^{-1}$ Mpc$^{-1}$ and cosmological
parameters $\Omega_{\rm M}$ = 0.273, $\Omega_\Lambda$ = 0.682 
were obtained from
the {\it Planck} mission data \citep{PC13}. The equivalent widths EW(H$\beta$)
were reduced to the rest frame.

The SDSS spectra were obtained with a small aperture of 3\arcsec\ in diameter. 
To derive integrated characteristics of the galaxies from their spectra and to 
make a comparison with photometric data more accurate, \citet{I14} 
corrected spectroscopic data for the aperture using the relation 
2.5$^{r(3\arcsec)-r}$, where $r$ and $r$(3\arcsec) are the 
SDSS $r$-band total magnitude
and the magnitude within the 3\arcsec\ spectroscopic aperture, respectively.

The SDSS spectra for some galaxies from the Herschel sample were also 
available. \citet{I14} used these data to derive the same parameters as those
for SDSS sample galaxies. In other cases, we used optical spectra
obtained with different telescopes to derive extinction, physical conditions
and element abundances. The H$\beta$ fluxes were obtained from SDSS spectra
only for three Herschel galaxies, HS 1222+3741, HS 1304+3529, and 
HS 1330+3651. For other galaxies, we adopted the observed H$\beta$ fluxes
from the literature, which were obtained with the largest 
available apertures and corrected them for the extinction derived from the 
optical spectra. References on the optical spectra and H$\beta$ fluxes are
given in Table \ref{tab0}.

\subsection{Stellar masses}

\citet{I14} derived galaxy stellar masses by modelling the galaxy spectral 
energy distributions (SED) in the optical range for each galaxy from the 
SDSS sample. They took into account both the stellar and ionised gas emission.

The method is based on fitting a series of model SEDs to the observed one 
and finding the best
fit. The fit, described in more detail by \citet{G06,G07} and \citet{I11a,I14}, 
was performed for each SDSS spectrum over the whole
observed spectral range of $\lambda$$\lambda$3900--9200\AA.
As each SED is the sum of both stellar and ionised gas emission,
its shape depends on the relative contribution of these two
components. The contribution of gaseous 
emission relative to stellar emission can be parameterized by the equivalent
width EW(H$\beta$) of the H$\beta$ emission line. The shape of the spectrum
depends also on reddening and the star-formation history of the galaxy.

\citet{I14} approximated the star-formation history in each galaxy by a recent 
short burst with age $<$ 10 Myr, which accounts 
for the young stellar population, and a prior continuous star formation 
responsible for the older stars with age $\geq$ 10 Myr.
The contribution of each stellar population to the SED was parameterized by the
varying ratio of the masses of the young to old stellar populations, 
$M_{\rm y}$/$M_{\rm o}$. Then the total stellar mass is equal to
$M_*$= $M_{\rm y}$ + $M_{\rm o}$. We adopted $M_*$ and $M_{\rm y}$, calculated 
by \citet{I14} for SDSS sample galaxies, where details of the SED fitting can 
be found. 

For the galaxies from the Herschel sample, 
we derive stellar masses, using the same method of SED fitting and spectra 
as cited in Table \ref{tab0}.

\subsection{Integrated dust characteristics}

To derive integrated dust characteristics in our SDSS and Herschel
samples we use the modified blackbody fitting technique, similar to that
used by \citet{RR13}. 

\citet{RR13} fitted the Herschel data by adopting a single
dust temperature $T$. This is sufficient to fit far-infrared
fluxes and to derive the mass of the cold dust. However, there is evidence 
for warm and hot dust in star-forming dwarf galaxies 
\citep[e.g. ][]{G11,I11b,I14,H13}. 

Therefore, we adopted a three-component
model with the temperatures $T_{\rm cd}$, $T_{\rm wd}$, and $T_{\rm hd}$ to
fit dust emission in the wavelength range $\sim$ 3 - 500 $\mu$m covered
by WISE, Spitzer, and Herschel observations:
%\begin{equation}
%F_\nu=\sum_{i=1}^n \left(\frac{a_i\nu^{3+\beta}}{\exp(h\nu/kT_i)-1}\right),
%\label{Fnu}
%\end{equation}
\begin{eqnarray}
F_\nu&=&\frac{a_{\rm cd}\nu^{3+\beta}}{\exp(h\nu/kT_{\rm cd})-1}+\frac{a_{\rm wd}\nu^{3+\beta}}{\exp(h\nu/kT_{\rm wd})-1}\\ \nonumber
     &+&\frac{a_{\rm hd}\nu^{3+\beta}}{\exp(h\nu/kT_{\rm hd})-1},
\label{Fnu}
\end{eqnarray}
where $h$ and $k$ are Planck and Boltzman constants, respectively,
$\beta$ is the emissivity index, 
$a_{\rm cd}$, and $a_{\rm wd}$, and $a_{\rm hd}$ are scaling coefficients.

The cold dust component with the temperature $T_{\rm cd}$ contributes mainly at
wavelengths $\geq$ 70 $\mu$m covered by Spitzer/{\it MIPS} 70$\mu$m and 
160$\mu$m and Herschel 
observations. The warm dust component with the temperature $T_{\rm wd}$
strongly
contributes in the wavelength range $\sim$ 10 - 50$\mu$m covered mainly
by the WISE 12$\mu$m and 22 $\mu$m, Spitzer/{\it MIPS} 24$\mu$m and
Spitzer/{\it IRS} observations. Finally, the hot dust component with the
temperature $T_{\rm hd}$ contributes in the wavelength range $<$ 10 $\mu$m,
which is covered by the WISE 3.4$\mu$m and 4.6 $\mu$m, and Spitzer/IRAC
observations. In reality, there is a gradient in dust temperature.
However, we show later that a three-component model is sufficient
to fit dust emission in the entire mid- and far-infrared ranges.

We use Monte Carlo simulations to derive $T_{\rm cd}$, $T_{\rm wd}$, and 
$T_{\rm hd}$, and scaling coefficients $a_{\rm cd}$, $a_{\rm wd}$, and $a_{\rm hd}$
by minimization of $\chi^2$:
\begin{equation}
\chi^2 = \sum_{j=1}^m \frac{[F_j({\rm obs})-F_j({\rm mod})]^2}{\sigma_j^2({\rm obs)}},
\label{chi2}
\end{equation}
where $F_j$(obs) and $F_j$(mod) are observed and modelled fluxes, 
$\sigma_j$ is the uncertainty of the observed flux, and $m$ is the number 
of observed fluxes.

The 1$\sigma$ errors (68.3\% confidence level) of $T_{\rm cd}$, $T_{\rm wd}$, 
$T_{\rm hd}$, $a_{\rm cd}$, $a_{\rm wd}$, and $a_{\rm hd}$ are estimated by adopting 
$\Delta$$\chi^2$ = $\chi^2$ -- $\chi^2_{\rm min}$ = 7.04 for six degrees of 
freedom. For some objects, we
considered SED modelling with a two-component dust, excluding the hot component.
In this case, $\Delta$$\chi^2$ = 4.72 for four degrees of freedom. Derived
errors of temperatures and scaling coefficients were then propagated to
obtain errors for dust masses and luminosities.

The dust masses $M_{\rm cd}$, $M_{\rm wd}$, and $M_{\rm hd}$ of each component 
are given by
\begin{equation}
%F_\nu (\lambda) = \frac{M_{\rm dust}\kappa (\lambda_0)}{D^2}
%\left(\frac{\lambda}{\lambda_0}\right)^{-\beta } B_\nu (\lambda ,T), 
M_{\rm dust}=\frac{D^2}{\kappa(\lambda_0)}
\left(\frac{\lambda}{\lambda_0}\right)^\beta
\frac{F_\nu(\lambda)}{B_\nu(\lambda,T_{\rm dust})}, \label{Mdust}
\end{equation}
where $\lambda_0$ = 100 $\mu$m, 
$\kappa$($\lambda_0$) = 34.7 cm$^2$ g$^{-1}$ is the dust emissivity 
cross section per unit mass at 100 $\mu$m for the best fit SMC 
(Small Magellanic Cloud)
dust \citep{WD01}, $D$ is the distance to the galaxy, $T_{\rm dust}$ =
$T_{\rm cd}$, $T_{\rm wd}$ or $T_{\rm hd}$ is the dust temperature,
$B_\nu$($\lambda$,$T_{\rm dust}$) is the Planck function, and $F_\nu$($\lambda$)
is the monochromatic flux of dust emission.

\citet{RR13} varied the emissivity index $\beta$ from 0 to 2.5 to
achieve the best agreement between the observed and modelled cold dust
emission.  At wavelengths $\ga$ 20$\mu$m, the dust absorption
coefficient is fairly well approximated by a power law $\nu^\beta$ fit
with $\beta$ = 2.0 \citep{D03,K08}, and as this is the range where most
of the energy is emitted, we use this value to calculate the dust
luminosity of the cold component from

\begin{equation}
L(T_{\rm dust}) = 4\pi M_{\rm dust}\kappa (\lambda_0)
\int \left(\frac{\lambda}{\lambda_0}\right)^{-\beta } B (\lambda ,T_{\rm dust})
d\lambda. \label{Ldust}
\end{equation}

At shorter wavelengths, $\beta$ is smaller and more variable, but, for
simplicity, we use the same $\beta$ also for the warm and hot
component. Nevertheless we also discuss the expected changes 
if $\beta$ is lower.

The photometric WISE and Spitzer fluxes at $\la$ 15 $\mu$m prior 
to fitting were corrected for contributions from the stellar and gaseous 
(continuum and emission lines) emission. 
The stellar and gaseous continuum can 
be important at $\lambda$ $\la$ 5 $\mu$m. We subtract it by
using extrapolations of SED fits obtained
from optical spectra. However, if observed WISE $\lambda$3.4$\mu$m,
$\lambda$4.6$\mu$m fluxes and Spitzer $\lambda$3.6$\mu$m,
$\lambda$4.5$\mu$m fluxes are smaller than the extrapolated fluxes, we excluded
them from the hot dust emission fitting. The stellar and gaseous continuum
is negligible at longer wavelengths and does not affect the determination
of warm and cold dust parameters.

%%%%%%%%%%%%%%%%%%%%%%%%%%%%%%%%%%%%%%%%%%%%%%%%
%    Fig.1 
%%%%%%%%%%%%%%%%%%%%%%%%%%%%%%%%%%%%%%%%%%%%%%%%
\begin{figure}%[t]
\hbox{
\includegraphics[angle=-90,width=0.98\linewidth]{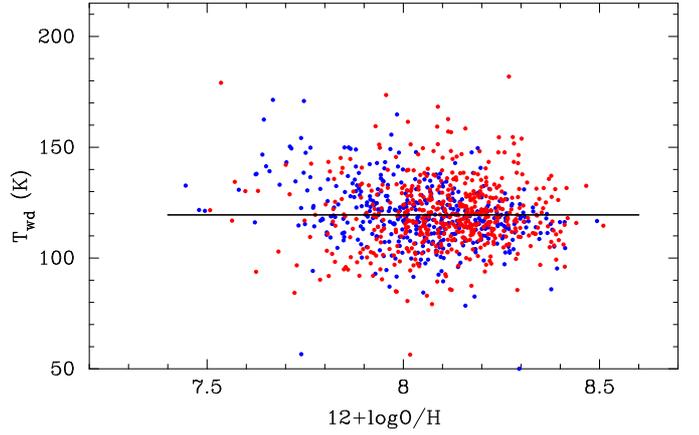}
}
%\hspace{0.2cm}\includegraphics[angle=-90,width=0.48\linewidth]{alhb_tmf.ps}}
%\hspace{0.2cm}\includegraphics[angle=-90,width=0.48\linewidth]{o_thf.ps}}
%\vspace{0.2cm}
%\hbox{
%\includegraphics[angle=-90,width=0.49\linewidth]{altot_thf.ps}
%\hspace{0.2cm}\includegraphics[angle=-90,width=0.49\linewidth]{alhb_thf.ps}}
\caption{Dependence of the warm dust temperature $T_{\rm wd}$ 
on the oxygen abundance 12+logO/H in $\sim$1000 galaxies from the SDSS sample.
Temperatures are calculated from the WISE fluxes at 12 $\mu$m
and 22 $\mu$m for galaxies with EW(H$\beta$) $\geq$50\AA\ (red filled
circles) and EW(H$\beta$) $<$50\AA\ (blue filled
circles). The solid line corresponds to average $T_{\rm wd}$.
%calculated with $\beta$ = 2.
%b) Same as in (a) but for the hot dust temperature $T_{\rm hd}$ 
%calculated from the WISE fluxes at 3.4 $\mu$m and 4.6 $\mu$m.
%Only galaxies with red colours 3.4$\mu$m -- 4.6$\mu$m $\geq$ 1.5 mag 
%are considered.
}
\label{fig1}
\end{figure}
%%%%%%%%%%%%%%%%%%%%%%%%%%%%%%%%%%%%%%%%%%%%%%%%%

%%%%%%%%%%%%%%%%%%%%%%%%%%%%%%%%%%%%%%%%%%%%%%%%
%    Fig.2 
%%%%%%%%%%%%%%%%%%%%%%%%%%%%%%%%%%%%%%%%%%%%%%%%
\begin{figure*}
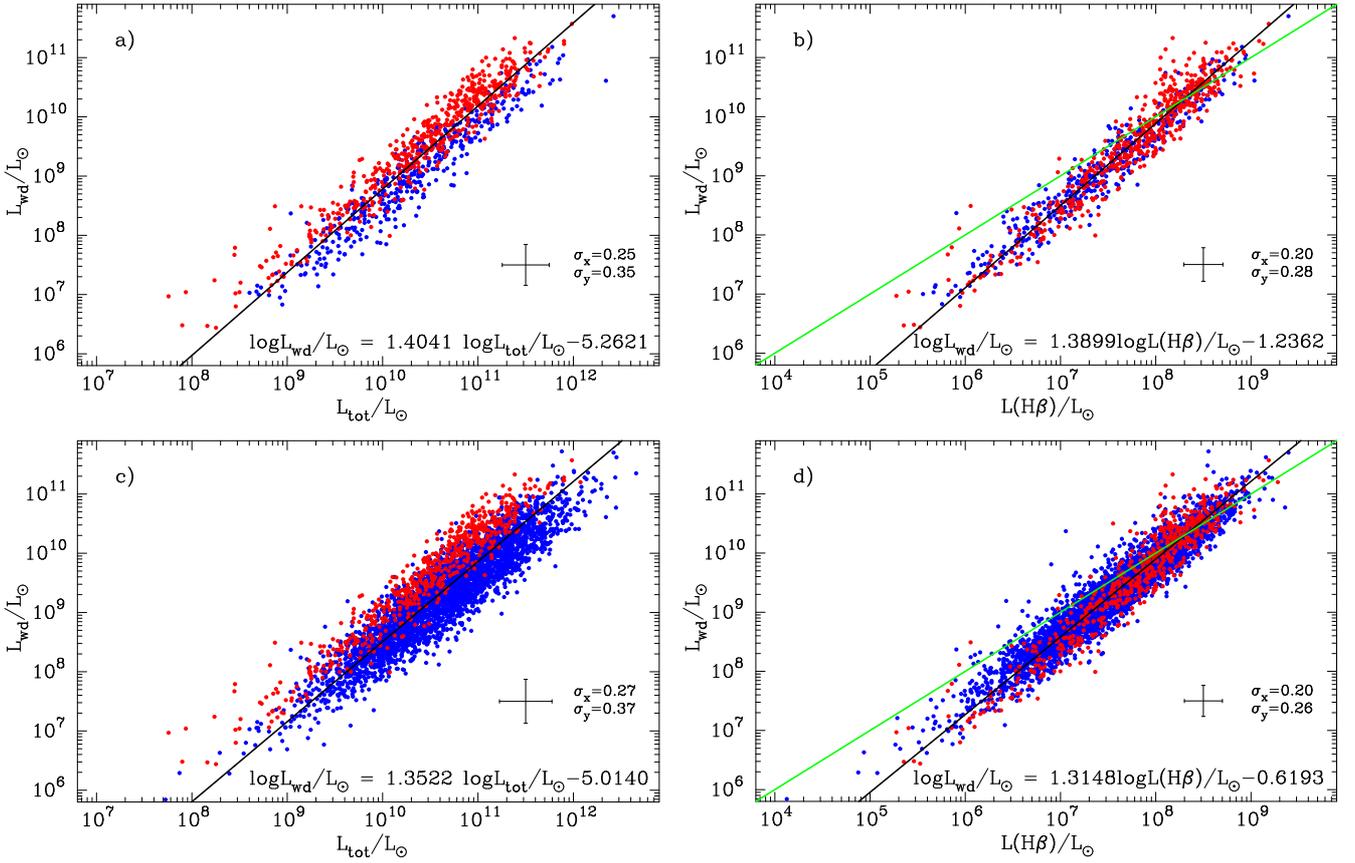
%[t]
%\centering{
\hbox{
\includegraphics[angle=-90,width=0.47\linewidth]{altot_alwarmdust.ps}
\hspace{0.2cm}\includegraphics[angle=-90,width=0.47\linewidth]{alhb_alwarmdust.ps}}
\vspace{0.2cm}
\hbox{
\includegraphics[angle=-90,width=0.47\linewidth]{altot_alwarmdust_all.ps}
\hspace{0.2cm}\includegraphics[angle=-90,width=0.47\linewidth]{alhb_alwarmdust_all.ps}}
%\hbox{
%\includegraphics[angle=-90,width=0.47\linewidth]{altot_alhotdust.ps}
%\hspace{0.2cm}\includegraphics[angle=-90,width=0.47\linewidth]{alhb_alhotdust.ps}}
\caption{Dependence of the warm dust luminosity $L_{\rm wd}$ of the sample
of $\sim$ 1000 SDSS compact galaxies with reliably derived oxygen abundance
on (a) total luminosity $L_{\rm tot}$ within 0.1 -- 22 $\mu$m 
and (b) H$\beta$ luminosity $
L$(H$\beta$). Red and blue filled circles are for galaxies with
EW(H$\beta$) $\geq$50\AA\ and EW(H$\beta$) $<$50\AA, respectively. 
(c) and (d) Same as in (a) and (b) but for the entire sample of 
$\sim$ 4000 SDSS compact star-forming galaxies. In all panels, 
black solid lines are fits to the data where 
$L_{\rm wd}$ is calculated by adopting $\beta$ = 2.
Green solid lines in (b) and (d) are one-to-one relations between warm dust 
luminosities and
luminosities of ionising radiation $L_{\rm ion}$ $\sim$ 100$L$(H$\beta$).}
\label{fig2}
\end{figure*}
%%%%%%%%%%%%%%%%%%%%%%%%%%%%%%%%%%%%%%%%%%%%%%%%%

%%%%%%%%%%%%%%%%%%%%%%%%%%%%%%%%%%%%%%%%%%%%%%%%
%    Fig.3
%%%%%%%%%%%%%%%%%%%%%%%%%%%%%%%%%%%%%%%%%%%%%%%%
%\begin{figure*}%[t]
%\hbox{
%\includegraphics[angle=-90,width=0.48\linewidth]{altot_mdust.ps}
%\hspace{0.2cm}\includegraphics[angle=-90,width=0.48\linewidth]{alhb_mdust.ps}}
%\caption{Dependence of the warm dust mass $M_{\rm wd}$ of the sample
%of $\sim$ 1000 SDSS compact galaxies with the reliably derived oxygen abundance
%on (a) total luminosity $L_{\rm tot}$ and (b) H$\beta$ luminosity $
%L$(H$\beta$). Red and blue filled circles are for galaxies with
%EW(H$\beta$) $\geq$50\AA\ and EW(H$\beta$) $<$50\AA, respectively. 
%In both panels, 
%solid, dashed and dotted lines are fits to the data where 
%$M_{\rm wd}$ is calculated adopting $\beta$ = 2, 1, and 0, respectively,
%and relations in right bottom corners are for the case with $\beta$ = 2.}
%\label{fig3}
%\end{figure*}
%%%%%%%%%%%%%%%%%%%%%%%%%%%%%%%%%%%%%%%%%%%%%%%%%

%%%%%%%%%%%%%%%%%%%%%%%%%%%%%%%%%%%%%%%%%%%%%%%%
%    Fig.3  z vs w1-w2
%%%%%%%%%%%%%%%%%%%%%%%%%%%%%%%%%%%%%%%%%%%%%%%%
\begin{figure*}
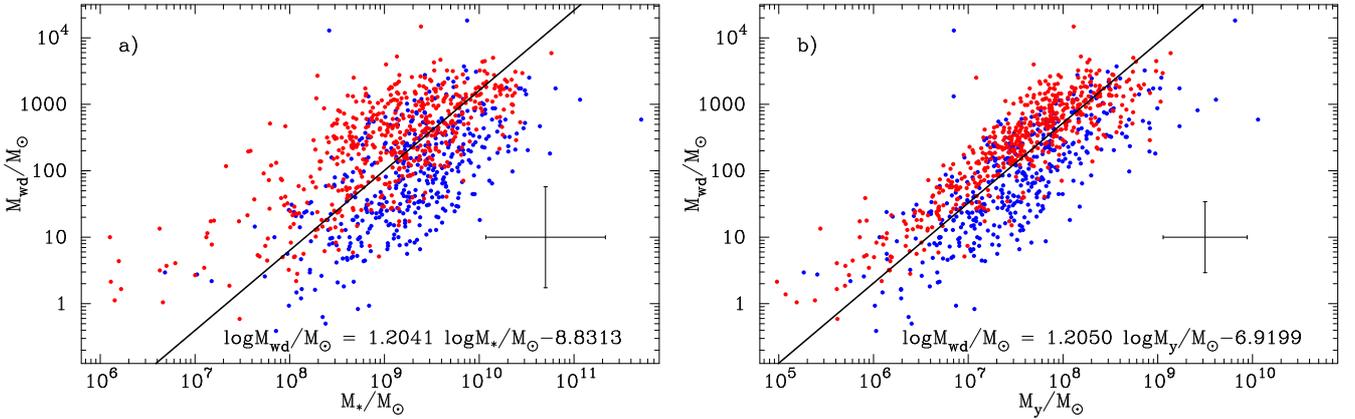
%[t]
\hbox{
\includegraphics[angle=-90,width=0.47\linewidth]{mtot_mdust.ps}
\hspace{0.2cm}\includegraphics[angle=-90,width=0.47\linewidth]{myoung_mdust.ps}}
\caption{Dependence of the warm dust mass $M_{\rm wd}$ of the sample
of $\sim$ 1000 SDSS compact galaxies with reliably derived oxygen abundance
on (a) total galaxy stellar mass $M_{\rm *}$ and (b) mass of the young stellar
population $M_{\rm y}$. Red and blue filled circles are for galaxies with
EW(H$\beta$) $\geq$50\AA\ and EW(H$\beta$) $<$50\AA, respectively. 
In both panels, solid lines are fits to the data where 
$M_{\rm wd}$ is calculated adopting $\beta$ = 2.}
\label{fig3}
\end{figure*}

%%%%%%%%%%%%%%%%%%%%%%%%%%%%%%%%%%%%%%%%%%%%%%%%
%    Fig.4  z vs w1-w2
%%%%%%%%%%%%%%%%%%%%%%%%%%%%%%%%%%%%%%%%%%%%%%%%
\begin{figure*}
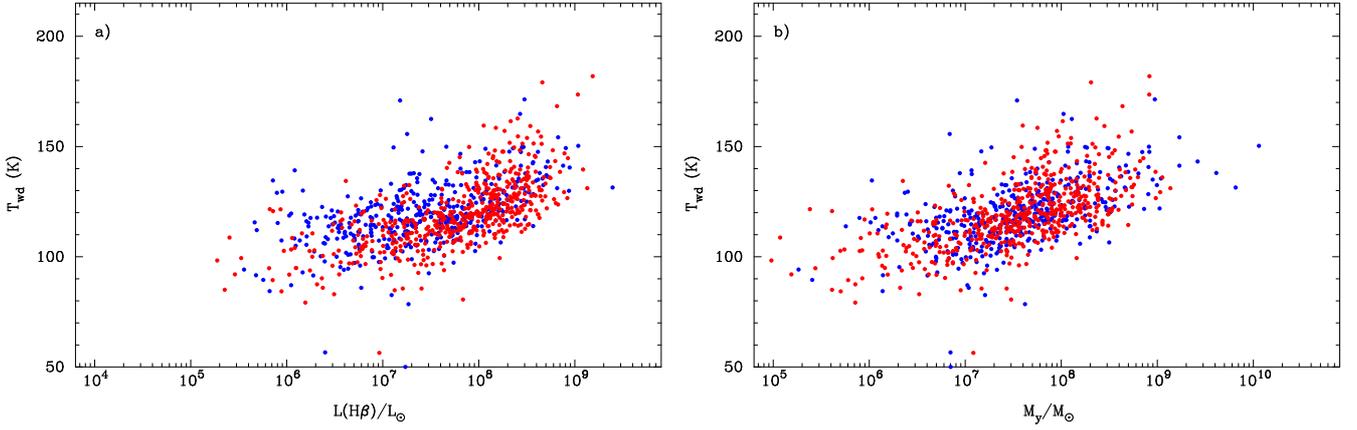
%[t]
\hbox{
\includegraphics[angle=-90,width=0.47\linewidth]{alhb_tmf.ps}
\hspace{0.2cm}\includegraphics[angle=-90,width=0.47\linewidth]{myoung_tmf.ps}}
\caption{Dependence of the warm dust temperature $T_{\rm wd}$ of the sample
of $\sim$ 1000 SDSS compact galaxies with a reliably derived oxygen abundance
on (a) H$\beta$ luminosity $L$(H$\beta$) and (b) mass of the young stellar
population $M_{\rm y}$. Red and blue filled circles are for galaxies with
EW(H$\beta$) $\geq$50\AA\ and EW(H$\beta$) $<$50\AA, respectively. 
In both panels, $T_{\rm wd}$ is calculated adopting $\beta$ = 2.}
\label{fig4}
\end{figure*}

%%%%%%%%%%%%%%%%%%%%%%%%%%%%%%%%%%%%%%%%%%%%%%%%
%    Fig.5  z vs w1-w2
%%%%%%%%%%%%%%%%%%%%%%%%%%%%%%%%%%%%%%%%%%%%%%%%
\begin{figure*}
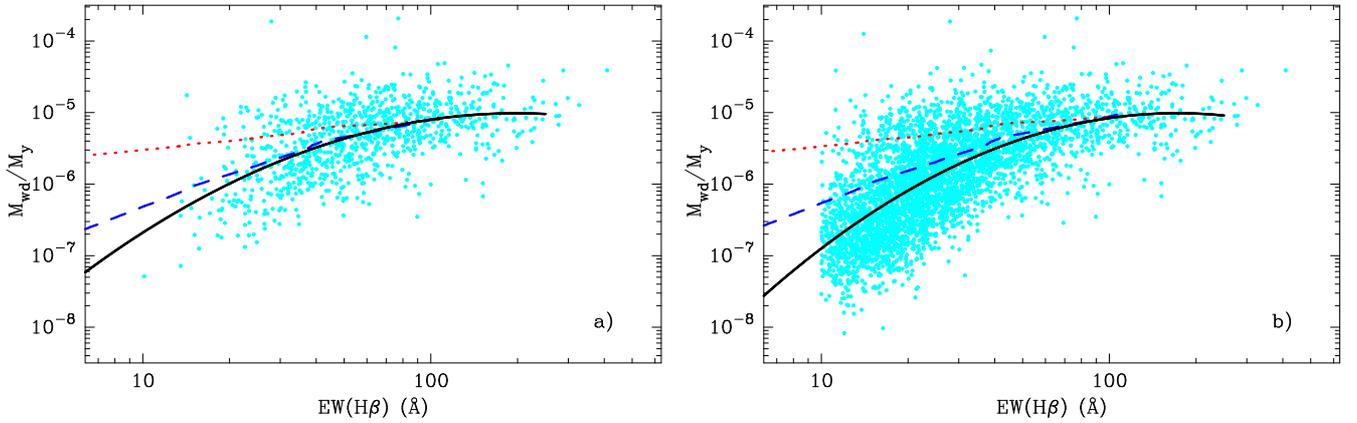
%[t]
\hbox{
\includegraphics[angle=-90,width=0.47\linewidth]{ewhb_mdust_myoung_1.ps}
\hspace{0.2cm}\includegraphics[angle=-90,width=0.47\linewidth]{ewhb_mdust_myoung_all_1.ps}}
%\hbox{
%\includegraphics[angle=-90,width=0.47\linewidth]{ewhb_mdust_mtot.ps}
%\hspace{0.2cm}\includegraphics[angle=-90,width=0.47\linewidth]{ewhb_mdust_mtot_all.ps}}
\caption{a) Dependence of the warm dust mass $M_{\rm wd}$ to the young
stellar population mass $M_{\rm y}$ ratio on the H$\beta$ 
equivalent width EW(H$\beta$) for the sample
of $\sim$ 1000 SDSS compact galaxies with a reliably derived oxygen abundance.
b) Same as in (a) but for the entire sample of $\sim$4000 SDSS compact 
star-forming galaxies. 
%Red and blue filled circles are for galaxies with
%EW(H$\beta$) $\geq$50\AA\ and EW(H$\beta$) $<$50\AA, respectively. 
In both panels, the warm dust mass $M_{\rm wd}$ is calculated by adopting
$\beta$ = 2. Solid lines are quadratic fits to the data, while 
dashed and dotted lines correspond to $M_{\rm wd}$/$M_{\rm y}$ ratios,
which are proportional to the luminosity $L_{\rm ion}$ of ionising
radiation and total luminosity $L_{\rm tot}$, respectively.
%c) Same as in (a) but for $M_{\rm wd}$/$M_{\rm *}$ ratio. d) Same as in (b) 
%but for $M_{\rm wd}$/$M_{\rm *}$ ratio.
}
\label{fig5}
\end{figure*}

The grid of CLOUDY spherical ionisation-bounded H~{\sc ii} region
models calculated in a wide range of input parameters 
\citep[version c13.01, ][]{F98,F13} is used to predict and subtract 
the strongest mid-infrared emission lines,
[Ar {\sc iii}] $\lambda$9.0 $\mu$m, [S {\sc iv}] $\lambda$10.51 $\mu$m, and
[Ne {\sc iii}] $\lambda$15.55 $\mu$m at their respective redshifted wavelengths.
The range of parameters and a grid are described by \citet{I13}.
The input CLOUDY chemical abundances were derived from the optical
spectra of each object. 
The equivalent widths of MIR emission lines ($\la$ 0.1 $\mu$m) are 
small as compared to the widths of the WISE photometric bands
($\sim$ 10 $\mu$m); therefore,
corrections for emission lines do not exceed $\sim$ 1\%.
The CLOUDY models also predict fluxes of PAH lines at $\lambda$3.3 $\mu$m,
$\lambda$6.2 $\mu$m, $\lambda$7.9 $\mu$m, $\lambda$11.3 $\mu$m, 
$\lambda$11.8 $\mu$m, and $\lambda$13.3 $\mu$m.
However, we do not subtract PAH emission from the observed fluxes
due to the complexity of PAH features fitting. Our inspection of 
the Spitzer/{\it IRS} spectra shows that PAH emission is visible only
in the six highest-metallicity galaxies from our Herschel sample
with 12+logO/H $\geq$8.1. Neglecting corrections for PAH emission would
overestimate the temperature of the hot dust component by at most 10\%. 
Parameters of the warm and cold components will not be changed.

Since only SDSS optical, 2MASS  NIR and WISE MIR data are available for 
the most of the SDSS sample galaxies, we cannot fit the cold dust component.
Therefore, for this sample, we adopted a one-component dust model to derive
characteristics of the warm dust from the flux ratio at $\lambda$12 $\mu$m and
$\lambda$22 $\mu$m. For a small number of 58 SDSS galaxies with red 
3.4$\mu$m -- 4.6$\mu$m colours of $\geq$ 1.5 mag, implying the presence of hot
dust, we also adopted a two-component dust model to derive characteristics of 
the hot and warm dust from the $\lambda$3.4$\mu$m/$\lambda$4.6$\mu$m and 
$\lambda$12$\mu$m/$\lambda$22$\mu$m flux ratios, respectively.

For the Herschel sample we use available data in the optical, 
NIR, MIR, FIR, sub-mm and mm ranges to fit the observed data with the
three-component dust model. However, the 
two-component dust model is sufficient for six galaxies. We note that 
we discuss the parameters of only cold and warm dust emission in all cases 
below. This is because hot dust emission may not be in 
thermal equilibrium.
Therefore, its characteristics, such as apparent temperature and mass, may 
be misleading and do not correctly represent its true parameters.

\section{Results \label{results}}

\subsection{SDSS sample of compact galaxies}

We use WISE 12$\mu$m and 22$\mu$m fluxes to derive the temperatures
$T_{\rm wd}$, the luminosities $L_{\rm wd}$, and 
the masses $M_{\rm wd}$ of the warm dust.
%Additionally, for SDSS galaxies with red colours 3.4$\mu$m--4.6$\mu$m $\geq$ 
%1.5 mag we derive the temperatures $T_{\rm hd}$, the luminosities $L_{\rm hd}$,
%and the masses $M_{\rm hd}$ of hot dust from the WISE 3.4$\mu$m/4.6$\mu$m
%flux ratio. The selection by colour is chosen because 
%red 3.4$\mu$m--4.6$\mu$m colours indicate the presence of hot dust
%\citep{G11,I11b,I14}. 

In Fig. \ref{fig1}, we show the dependence of the warm dust temperature
$T_{\rm wd}$ on the oxygen abundance 12 +log O/H that is calculated with the 
emissivity index $\beta$ =2.0. Only $\sim$1000 galaxies with 
reliably derived oxygen abundances are shown in the Figure. 
Galaxies with the H$\beta$ equivalent width
EW(H$\beta$) $\geq$50\AA\ and $<$50\AA\ are also represented. 
Most of the galaxies are spread in the $T_{\rm wd}$ 
interval of $\sim$ 80 -- 150K with an average value of $\sim$120K. 
The average temperature would be higher by $\sim$ 20K if
$\beta$ = 1.0 is adopted.

No clear dependence of dust temperature on 12+logO/H is found, implying that 
the metallicity is not a factor, which regulates dust emission. We also do not
find differences between warm dust temperatures in galaxies with 
high-excitation and low-excitation H~{\sc ii} regions.
%We can not conclude the same for the hot dust temperature because statistics 
%are poor and all but one galaxy are sources with high-excitation H~{\sc ii}
%regions (Fig. \ref{fig1}b). 

Figures \ref{fig2}a and \ref{fig2}b respectively
show dependencies of warm dust luminosities $L_{\rm wd}$ on total
galaxy luminosities $L_{\rm tot}$ and on H$\beta$ luminosity $L$(H$\beta$)
for $\sim$1000 compact galaxies with reliable oxygen abundances. 
The total luminosity $L_{\rm tot}$ is calculated in the wavelength 
range 0.1 -- 22 $\mu$m \citep{I13}.
Linear most likelihood fits of the relations are shown by solid lines.

We find
that the relation in Fig. \ref{fig2}b is tighter than that in Fig. \ref{fig2}a.
The difference is more obvious from the comparison of Fig. \ref{fig2}c and
Fig. \ref{fig2}d, where we show the entire SDSS sample of $\sim$4000 compact
star-forming galaxies. This difference indicates that warm dust is heated 
by the UV radiation of star-forming regions with a significant fraction of
ionising stellar radiation. Thus much of the warm dust is likely associated
with H~{\sc ii} regions and neutral gas clouds
surrounding these H~{\sc ii} regions.

This conclusion is supported by the result that there is no offset between
galaxies with high and low EW(H$\beta$) (Figs. \ref{fig2}b,d),
implying that the warm dust luminosity
is determined by the $L$(H$\beta$) luminosity which in turn is proportional
to the luminosity of ionising radiation. The relations on the total
luminosities, which include both ionising and non-ionising radiation, are
broader with clear separation of galaxies with high and low EW(H$\beta$)
(Figs. \ref{fig2}a,c).

The dependence of the warm dust mass 
$M_{\rm wd}$, calculated with $\beta$ = 2.0,
on total stellar mass $M_*$ is shown in Fig. \ref{fig3}a
for galaxies with high and low EW(H$\beta$), respectively.
The warm dust mass would be increased
by a factor of $\sim$ 2 if $\beta$ = 1.0 is adopted.
At a variance with the $L_{\rm wd}$ - $L_{\rm tot}$ relation, the correlation 
$M_{\rm wd}$ - $M_*$ is much weaker. This can be due to the presence of a
faint old stellar population, not participating in the heating of warm dust. 
Its contribution to the stellar mass can be high, while it does not contribute
much to the total luminosity because of the steeper than linear relation
between stellar mass and luminosity.

The dependence of $M_{\rm wd}$ on the mass of the young stellar population
$M_{\rm y}$ is much tighter (Fig. \ref{fig3}b) supporting the conclusion
that warm dust in SDSS compact dwarf galaxies is 
heated to a large extent by the radiation 
from young stars. Similarly to Fig. \ref{fig3}a, the warm dust mass is
increased by a factor of $\sim$ 2, if an emissivity
index $\beta$ of 1.0 is adopted.
However, a clear offset is present in both panels of Fig. \ref{fig3} - 
the mass of warm dust in galaxies with lower EW(H$\beta$)
is lower than that in galaxies with high EW(H$\beta$).

Relations $L_{\rm wd}$ vs. $L$(H$\beta$) (Fig. \ref{fig2}b) and
$M_{\rm wd}$ vs. $M_{\rm y}$ (Fig. \ref{fig3}b) are steeper than linear relations.
We may assume that the non-linearity of the $L_{\rm wd}$ vs. $L$(H$\beta$)
relation is due to the result that the warm dust in more
luminous and more massive starbursts is warmer and is thus characterised
by higher emissivity (Fig. \ref{fig4}a). However, in this case, one would
expect a shallower than linear relation $M_{\rm wd}$ vs. $M_{\rm y}$ because of the
positive correlation between $T_{\rm wd}$ and  $M_{\rm y}$ (Fig. \ref{fig4}b) 
although it is steeper. Probably, other factors such as the dust-to-gas mass 
ratios play a role in steepening relations in Figs. \ref{fig2}b and 
\ref{fig3}b. 
The dust-to-gas mass ratios are smaller for lower-metallicity galaxies, which
are also fainter and less massive. Possible deviations of warm dust emission 
from the thermal equilibrium may also play a role.
%, although we have no observational
%evidence for the existence of such deviations.

We produce dependencies of the $M_{\rm wd}$/$M_{\rm y}$ ratios on EW(H$\beta$) for
$\sim$1000 SDSS galaxies with best derived oxygen abundances and for 
the entire sample of $\sim$4000 SDSS compact galaxies in Fig. \ref{fig5}a 
and \ref{fig5}b.
The clear decrease of the mass ratio with decreasing EW(H$\beta$) or 
respective increasing
age is seen. It is more evident for the entire sample (Fig. \ref{fig5}b). 
This decrease is caused mainly by the decrease of $M_{\rm wd}$, because
$M_{\rm y}$ is almost not changed at EW(H$\beta$) $\geq$ 10\AA, corresponding
to instantaneous burst ages $<$ 10 Myr.

We fit the distributions for both samples by
quadratic maximum-likelihood regressions. For comparison,
we also show dependencies of $M_{\rm wd}$/$M_{\rm y}$ ratios on EW(H$\beta$), 
which are proportional to the luminosities of ionising radiation and to the 
total luminosities of the instantaneous burst 
\citep{SV98,L99}. It is seen that these dependencies are not as steep, but the 
dependencies, which are proportional to the luminosities of ionising radiation, 
are in closer agreement with the likelihood regressions.
The differences between the dependencies are
likely due to that the $M_{\rm wd}$/$M_{\rm y}$ ratios increase with the 
luminosity of ionising radiation more steeply than linearly.
Thus, this result and the tight correlation
between the luminosities $L_{\rm wd}$ and $L$(H$\beta$) suggest that
warm dust is associated with H~{\sc ii} regions
and is heated in significant part by ionising
radiation of hot O-stars beyond the Ly-limit at 912\AA.

\subsection{Herschel sample}

The original Herschel sample of 48 galaxies by \citet{RR13} is more 
heterogeneous as compared to the SDSS 
galaxies considered in the previous subsection and includes extended irregular
galaxies and BCDs. Large angular sizes
of some galaxies in the Herschel sample make a comparison of data 
obtained at different wavelengths uncertain as large aperture corrections
are needed. On the other hand, these are nearby objects, some of them are
bright, and most of them were extensively observed in different wavelength
ranges. Applying selection criteria discussed in Sect. \ref{Herschel}, we
reduced the sample to 28 compact galaxies.
We collect the available data for the reduced Herschel sample
from the UV-range to the radio-range using mainly databases of all-sky 
surveys. References to these observational data are given in Table \ref{tab1}.

\subsubsection{SED fits \label{SED}}

The SED fits of stellar and ionised gas emission at $\lambda$ $\leq$ 15 $\mu$m,
of dust emission at $\lambda$ $\sim$ 2 - 1000 $\mu$m and of free-free 
emission at $\lambda$~$\geq$~100~$\mu$m and the 
observed data are shown in Fig.~\ref{fig6}. We note that modelled fluxes 
in this Figure are redshifted to observed wavelengths and reddened adopting 
the extinction coefficient $C$(H$\beta$), which is derived from optical 
spectra and the reddening law by \citet{C89} with $R_V$ = 3.2 
(cf. Sect.~\ref{abund}).
This is done to make a comparison with observed fluxes.

First, the optical spectra were
used to derive line intensities and equivalent widths of emission lines,
physical conditions, redshifts and extinction. We produce
spectral energy distributions, 
which are the sum of stellar SED and ionised gas 
SED. These fits were extrapolated to the UV and IR
ranges and cover a wavelength range of 0.1 -- 15 $\mu$m. 
Observed fluxes in optical spectra prior to the 
fitting were corrected for aperture. The aperture
correction for nine compact galaxies with SDSS spectra was done comparing the
photometric $r$ magnitude of the galaxy with $r$ magnitude measured within
the spectroscopic slit of 3\arcsec\ in diameter. 
Typical aperture corrections for compact Herschel 
galaxies are of order $\sim$ 1.5 - 2.0. 
For other galaxies,
we scale the observed spectra by comparing H$\beta$ or H$\alpha$ line fluxes
inside the slit with those derived from narrow-band imaging and 
from spectra, which were obtained with large apertures. 
%These data are  available in the NED. 
%They can be higher by a factor of 
%more than ten for extended galaxies and are then much more uncertain.

Second, the extinction- and aperture-corrected flux of the H$\beta$ emission
line was used to calculate the SED of free-free emission in the sub-mm and 
radio ranges (at $\lambda$ $\geq$ 100$\mu$m in 
Fig. \ref{fig6}), according to \citet{CD86}.
%This SED was derived in all objects with available H$\beta$ or H$\alpha$ 
%fluxes.
% even if no optical spectrum is available.

Third, we find that the three-component model with cold, 
warm, and hot dust is generally required to fit the SED in the mid- and 
far-infrared ranges. For this, Spitzer/{\it IRS} 
spectra with varying wavelength ranges, but typically at 
$\lambda$5.5 -- 35$\mu$m, and available Herschel, Spitzer, and WISE 
photometric data, are shown by line and different symbols in Fig. \ref{fig6}. 
The two-component model without hot dust
is sufficient only in six galaxies
(HS~0052$+$2536, HS~0822+3542, I~Zw~18, Mrk~209, UGC~4483, and VII~Zw~403). 
The SED fits of individual dust components are shown
for hot, warm, and cold dust, respectively.
The total SED of dust emission is also shown, which additionally includes
the free-free emission component at $\lambda$ $\geq$ 100$\mu$m.
The wavelength at which dust emission 
and free-free emission are equal are also indicated. In all cases, 
a dust emissivity index $\beta$ of 2.0 is adopted.

\subsubsection{Comparison with \citet{RR13} results}

\renewcommand{\baselinestretch}{1.0}

 \begin{table*}
 \caption{Global parameters of galaxies from the {\sl Herschel} sample.}
 \label{tab1}
 \begin{tabular}{lrrcrrcrrccrr} \hline \hline
 &\multicolumn{8}{c}{Dust parameters}&12+ &    log &    log & \multicolumn{1}{c}{log}          \\ \cline{2-9}
 &\multicolumn{2}{c}{Temperature (K)}&&
\multicolumn{2}{c}{log Mass ($M_\odot$)}&&
\multicolumn{2}{c}{log Luminosity ($L_\odot$)}&logO/H&$L$(H$\beta$)$^{\rm a}$&$M_{\rm y}$&\multicolumn{1}{c}{$M_{\rm *}$}  \\      \cline{2-3}  \cline{5-6} \cline{8-9} 
Object &\multicolumn{1}{c}{cold}&\multicolumn{1}{c}{warm}&&
\multicolumn{1}{c}{cold}&\multicolumn{1}{c}{warm}&&
\multicolumn{1}{c}{cold}&\multicolumn{1}{c}{warm}&&($L_\odot$)&($M_\odot$)&\multicolumn{1}{c}{($M_\odot$)} \\ \hline
Haro 11        & 34$_{-2}^{+1}$~\,   &104$_{-9}^{+4}$~\,  &&  6.54$_{-0.03}^{+0.09}$&    3.73$_{-0.12}^{+0.28}$&&10.74$_{-0.12}^{+0.06}$&10.82$_{-0.23}^{+0.10}$&8.36&9.37& 9.45&10.51 \\
Haro 3         & 26$_{-1}^{+1}$~\,   & 93$_{-6}^{+6}$~\,  &&  5.89$_{-0.08}^{+0.03}$&    2.32$_{-0.13}^{+0.08}$&& 9.39$_{-0.04}^{+0.11}$& 9.13$_{-0.16}^{+0.16}$&8.28&6.74& 5.97& 7.76 \\
HS 0017$+$1055 & 38$_{-7}^{+12}$     &121$_{-4}^{+11}$    &&  4.11$_{-0.49}^{+0.42}$&    1.07$_{-0.29}^{+0.12}$&& 8.55$_{-0.55}^{+0.70}$& 8.56$_{-0.09}^{+0.22}$&7.63&6.96& 6.96& 7.22 \\
HS 0052$+$2536 & 36$_{-9}^{+1}$~\,   & 90$_{-1}^{+9}$~\,  &&  5.73$_{-0.04}^{+0.57}$&    2.85$_{-0.30}^{+0.02}$&&10.06$_{-0.71}^{+0.07}$& 9.58$_{-0.01}^{+0.25}$&8.04&7.90& 6.90& 8.83 \\
HS 0822$+$3542 & 33$_{-2}^{+5}$~\,   &107$_{-1}^{+12}$    &&  2.83$_{-0.47}^{+0.18}$& $-$0.92$_{-0.29}^{+0.03}$&& 6.94$_{-0.08}^{+0.04}$& 6.24$_{-0.02}^{+0.03}$&7.45&5.28& 4.60& 5.97 \\
HS 1222$+$3741 & 54$_{-12}^{+7}$     &111$_{-7}^{+11}$    &&  3.87$_{-0.43}^{+0.55}$&    1.70$_{-0.29}^{+0.25}$&& 9.25$_{-0.12}^{+0.09}$& 8.96$_{-0.08}^{+0.05}$&7.79&7.25& 6.87& 8.13 \\
HS 1304$+$3529 & 32$_{-3}^{+2}$~\,   & 72$_{-6}^{+6}$~\,  &&  4.96$_{-0.18}^{+0.20}$&    2.53$_{-0.29}^{+0.30}$&& 9.02$_{-0.07}^{+0.05}$& 8.68$_{-0.07}^{+0.06}$&7.93&7.24& 6.76& 8.19 \\
HS 1330$+$3651 & 27$_{-1}^{+3}$~\,   & 82$_{-5}^{+8}$~\,  &&  5.29$_{-0.18}^{+0.10}$&    2.01$_{-0.28}^{+0.16}$&& 8.89$_{-0.03}^{+0.05}$& 8.48$_{-0.05}^{+0.01}$&7.98&6.81& 7.42& 9.34 \\
I Zw 18        & 45$_{-1}^{+2}$~\,   &118$_{-3}^{+3}$~\,  &&  2.38$_{-0.11}^{+0.09}$& $-$0.67$_{-0.07}^{+0.09}$&& 7.31$_{-0.02}^{+0.02}$& 6.75$_{-0.02}^{+0.02}$&7.17&6.13& 5.94& 6.42 \\
II Zw 40       & 28$_{-1}^{+2}$~\,   &101$_{-6}^{+2}$~\,  &&  5.25$_{-0.11}^{+0.05}$&    1.87$_{-0.03}^{+0.19}$&& 8.95$_{-0.05}^{+0.18}$& 8.88$_{-0.17}^{+0.06}$&8.23&7.37& 7.93& 8.00 \\
Mrk 1089       & 25$_{-2}^{+3}$~\,   & 77$_{-5}^{+6}$~\,  &&  6.84$_{-0.25}^{+0.17}$&    3.56$_{-0.29}^{+0.21}$&&10.23$_{-0.03}^{+0.06}$& 9.88$_{-0.10}^{+0.03}$&8.10&7.93& 8.02& 8.89 \\
Mrk 1450       & 32$_{-1}^{+3}$~\,   & 98$_{-2}^{+5}$~\,  &&  4.01$_{-0.14}^{+0.04}$&    1.00$_{-0.20}^{+0.08}$&& 8.07$_{-0.03}^{+0.12}$& 7.95$_{-0.07}^{+0.03}$&7.84&6.48& 5.77& 7.55 \\
Mrk 153        & 33$_{-2}^{+2}$~\,   & 88$_{-1}^{+2}$~\,  &&  4.70$_{-0.11}^{+0.12}$&    1.61$_{-0.04}^{+0.02}$&& 8.77$_{-0.16}^{+0.14}$& 8.29$_{-0.01}^{+0.04}$&7.86&6.95& 7.84& 8.60 \\
Mrk 209        & 32$_{-2}^{+3}$~\,   & 96$_{-1}^{+9}$~\,  &&  3.11$_{-0.15}^{+0.13}$&    0.13$_{-0.30}^{+0.13}$&& 7.14$_{-0.05}^{+0.08}$& 7.01$_{-0.07}^{+0.01}$&7.74&5.83& 5.56& 5.83 \\
Mrk 930        & 29$_{-2}^{+3}$~\,   & 89$_{-5}^{+8}$~\,  &&  6.24$_{-0.22}^{+0.13}$&    2.99$_{-0.23}^{+0.19}$&& 9.98$_{-0.06}^{+0.04}$& 9.66$_{-0.02}^{+0.03}$&8.03&7.93& 7.78& 9.46 \\
NGC 1140       & 25$_{-1}^{+1}$~\,   & 80$_{-1}^{+3}$~\,  &&  5.97$_{-0.06}^{+0.04}$&    2.38$_{-0.12}^{+0.02}$&& 9.33$_{-0.01}^{+0.02}$& 8.79$_{-0.04}^{+0.01}$&8.38&7.12& 6.98& 8.72 \\
Pox 186        & 25$_{-6}^{+2}$~\,   & 80$_{-6}^{+9}$~\,  &&  3.85$_{-0.13}^{+0.59}$&    0.89$_{-0.28}^{+0.25}$&& 7.24$_{-0.11}^{+0.06}$& 7.29$_{-0.01}^{+0.01}$&7.70&5.65& 4.98& 6.33 \\
SBS 0335$-$052E& 57$_{-12}^{+7}$     &129$_{-8}^{+3}$~\,  &&  3.06$_{-0.16}^{+0.25}$&    1.37$_{-0.06}^{+0.19}$&& 8.61$_{-0.34}^{+0.14}$& 9.03$_{-0.01}^{+0.02}$&7.30&7.19& 6.78& 7.26 \\
SBS 1159$+$545 & 33$_{-8}^{+15}$     & 88$_{-1}^{+9}$~\,  &&  3.78$_{-0.70}^{+0.62}$&    1.34$_{-0.29}^{+0.04}$&& 7.85$_{-0.16}^{+0.31}$& 7.99$_{-0.01}^{+0.03}$&7.44&6.35& 5.76& 5.77 \\
SBS 1211$+$540 & 42$_{-7}^{+15}$     & 99$_{-9}^{+7}$~\,  &&  2.46$_{-0.45}^{+0.35}$& $-$0.48$_{-0.28}^{+0.30}$&& 7.22$_{-0.13}^{+0.31}$& 6.50$_{-0.10}^{+0.04}$&7.58&5.79& 5.67& 7.06 \\
SBS 1249$+$493 & 28$_{-6}^{+10}$     & 87$_{-6}^{+12}$    &&  5.10$_{-0.70}^{+0.62}$&    1.97$_{-0.30}^{+0.15}$&& 8.79$_{-0.04}^{+0.09}$& 8.60$_{-0.03}^{+0.04}$&7.68&6.90& 6.53& 7.41 \\
SBS 1415$+$437 & 37$_{-5}^{+5}$~\,   & 89$_{-6}^{+6}$~\,  &&  3.27$_{-0.22}^{+0.25}$&    0.59$_{-0.26}^{+0.24}$&& 7.65$_{-0.12}^{+0.09}$& 7.27$_{-0.09}^{+0.07}$&7.55&6.11& 5.35& 6.06 \\
SBS 1533$+$574 & 26$_{-1}^{+2}$~\,   & 91$_{-9}^{+9}$~\,  &&  5.34$_{-0.16}^{+0.04}$&    2.11$_{-0.29}^{+0.30}$&& 8.83$_{-0.02}^{+0.04}$& 8.87$_{-0.04}^{+0.04}$&8.05&6.70& 6.30& 8.08 \\
Tol 1214$-$277 & 37$_{-9}^{+17}$     & 87$_{-9}^{+11}$    &&  4.18$_{-0.65}^{+0.62}$&    1.93$_{-0.30}^{+0.28}$&& 8.56$_{-0.19}^{+0.32}$& 8.57$_{-0.01}^{+0.01}$&7.52&7.54& 6.94& 6.99 \\
UGC 4483       & 30$_{-5}^{+11}$     & 89$_{-3}^{+8}$~\,  &&  2.31$_{-0.70}^{+0.58}$& $-$1.17$_{-0.30}^{+0.10}$&& 6.21$_{-0.09}^{+0.09}$& 5.52$_{-0.07}^{+0.01}$&7.46&4.67& 4.46& 5.16 \\
UM 448         & 29$_{-1}^{+3}$~\,   & 87$_{-1}^{+8}$~\,  &&  6.84$_{-0.20}^{+0.10}$&    3.74$_{-0.24}^{+0.02}$&&10.65$_{-0.03}^{+0.07}$&10.37$_{-0.02}^{+0.02}$&8.32&8.23& 8.59&10.45 \\
UM 461         & 31$_{-1}^{+3}$~\,   & 88$_{-5}^{+1}$~\,  &&  3.57$_{-0.16}^{+0.10}$&    0.66$_{-0.02}^{+0.16}$&& 7.52$_{-0.01}^{+0.06}$& 7.33$_{-0.01}^{+0.02}$&7.73&5.83& 5.45& 6.94 \\
VII Zw 403     & 29$_{-1}^{+2}$~\,   & 78$_{-1}^{+7}$~\,  &&  3.33$_{-0.12}^{+0.07}$& $-$0.20$_{-0.25}^{+0.01}$&& 7.12$_{-0.04}^{+0.04}$& 6.15$_{-0.01}^{+0.03}$&7.66&5.22& 4.83& 5.98 \\
 \hline
 \end{tabular}
   
$^{\rm a}$extinction-corrected.
 \end{table*}

In Fig. \ref{fig6}, we use fits  to derive temperatures, luminosities, and
masses of cold, warm, and hot dust components in 28 compact galaxies 
%out of 48 galaxies 
from the Herschel sample  by \citet{RR13}. 

Derived temperatures, luminosities, and masses of cold and warm dust 
components are shown in Table \ref{tab1}, and they can be
compared to the properties of the cold dust derived by \citet{RR13}.
We do not show parameters for the hot dust component, because 
the equilibrium conditions for this component may not be valid.
Additionally, the Table shows oxygen abundances,
extinction-corrected H$\beta$ luminosities, masses of the young stellar 
population $M_{\rm y}$ and total stellar masses $M_*$.

We note that \citet{RR13} did not derive parameters of the cold dust 
in objects, which were not detected at 160$\mu$m and longer wavelengths. 
This refers to seven galaxies, HS~0822+3542, HS~1222+3741, I~Zw~18, 
SBS~1159+545, SBS~1249+493, Tol~1214$-$277, and UGC~4483, out of the 28 
galaxies from our sample. 
%Therefore, we compare properties of dust in 21 galaxies 
%derived by \citet{RR13} and in this paper. 
We also note, that we had to correct the luminosity of cold dust
in SBS~0335$-$052E, mistakenly given by \citet{RR13} (their Table 4)
upwards by a factor of $\sim$ 100.

\setcounter{figure}{6}

%%%%%%%%%%%%%%%%%%%%%%%%%%%%%%%%%%%%%%%%%%%%%%%%%%%%%%%%%%%%%%%%
% Fig 7 comparison with Remy-Ruyer et al. (2013)
%%%%%%%%%%%%%%%%%%%%%%%%%%%%%%%%%%%%%%%%%%%%%%%%%%%%%%%%%%%%%%%%
\begin{figure}
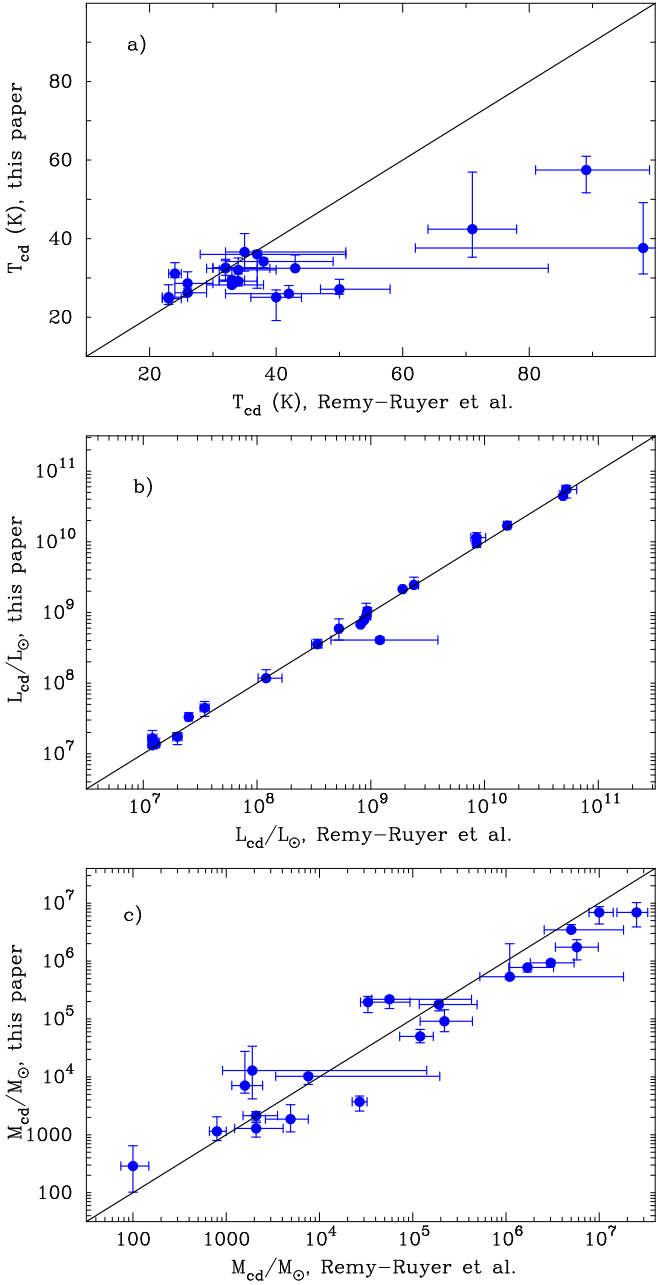

\hbox{
\includegraphics[width=5.5cm,angle=-90]{comp_t.ps}
}
\vspace{0.2cm}
\hbox{
\includegraphics[width=5.5cm,angle=-90]{comp_l.ps}
}
\vspace{0.2cm}
\hbox{
\includegraphics[width=5.5cm,angle=-90]{comp_m.ps}
}
%\psfig{figure=comp_t.ps,angle=-90,width=8.8cm,clip=}
%\vspace{0.1cm}
%\hspace*{0.0cm}\psfig{figure=comp_l.ps,angle=-90,width=8.8cm,clip=}
%\vspace{0.1cm}
%\hspace*{0.0cm}\psfig{figure=comp_m.ps,angle=-90,width=8.8cm,clip=}
%\hspace*{0.2cm}\psfig{figure=comp_t_beta.ps,angle=-90,width=4.4cm,clip=}
%}
\caption{The relation between the  (a) temperatures, (b) luminosities, and 
(c) masses of cold dust obtained for the Herschel sample in this paper 
and by \citet{RR13}. In all panels, solid lines mark a one to one 
correspondence.
%In the last panel of the figure the differences 
%between dust temperature obtained in this paper and the paper of 
%\citet{RR13} are plotted against parameter $\beta$ of modified 
%(blackbody) Planck's law.
}
\label{fig7}
\end{figure}
%%%%%%%%%%%%%%%%%%%%%%%%%%%%%%%%%%%%%%%%%%%%%%%%%%%%%%%%%%%%%%%%

In Fig. \ref{fig7}a, we show a comparison between the cold dust temperatures
in 21 galaxies, for which data are available in \citet{RR13} 
with those in this paper.
For most of the galaxies the agreement is good. 
However, there
are some outliers. These are objects, for which \citet{RR13} adopted
$\beta$ $\sim$0.0. The most deviant object is HS~0017+1055. We 
derive $T_{\rm cd}$ = 38$^{+12}_{-7}$K, using a three-component dust model.
Our fit reproduces all available WISE, Spitzer/{\it IRS} and
Herschel observations of this galaxy well, and the derived $T_{\rm cd}$ is
similar to that obtained for other galaxies (Table \ref{tab1}).
\citet{RR13} found $T_{\rm cd}$ = 98$^{+34}_{-36}$K, 
using a one-component dust model with $\beta$ = 0.0$^{+1.34}_{-0.00}$. 
%This difference is resulted
%in $\sim$ 10 times lower cold dust mass derived by \citet{RR13}.

On the other hand, the luminosities of the cold dust are
in good agreement (Fig. \ref{fig7}b). This is expected, because fits in both
cases reproduce the observed FIR fluxes. Derived masses of cold dust 
differ to a larger extent (Fig. \ref{fig7}c), despite the overall agreement 
between \citet{RR13} and our determinations. 
%The most deviant galaxies are
%those, for which \citet{RR13} obtained $\beta$ $\sim$ 0.0 
%(their Table 4) and thus $\sim$10 times lower masses of cold dust.

%%%%%%%%%%%%%%%%%%%%%%%%%%%%%%%%%%%%%%%%%%%%%%%%%%%%%%%%%%%%%%%%
% Fig 8 dust temperatures vs 12+logO/H
%%%%%%%%%%%%%%%%%%%%%%%%%%%%%%%%%%%%%%%%%%%%%%%%%%%%%%%%%%%%%%%%

\begin{figure}
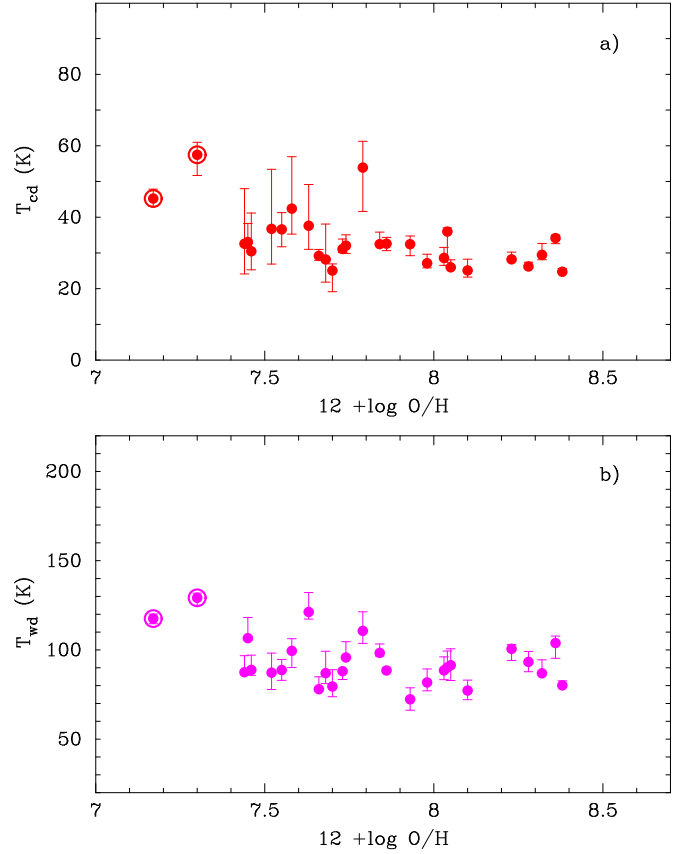

\hbox{
\includegraphics[width=5.5cm,angle=-90]{t_low_metal.ps}
}
\vspace{0.2cm}
\hbox{
\includegraphics[width=5.5cm,angle=-90]{t_w_metal.ps}
}
%\vspace{0.2cm}
%\hbox{
%\includegraphics[width=5.5cm,angle=-90]{t_h_metal.ps}
%}
%\hspace*{0.0cm}\psfig{figure=t_low_metal.ps,angle=-90,width=8.8cm,clip=}
%\vspace{0.1cm}
%\hspace*{0.0cm}\psfig{figure=t_w_metal.ps,angle=-90,width=8.8cm,clip=}
%\vspace{0.1cm}
%\hspace*{0.0cm}\psfig{figure=t_h_metal.ps,angle=-90,width=8.8cm,clip=}
%}
\caption{Dependencies of (a) cold and (b) warm dust temperatures for 
the Herschel sample on oxygen abundances 12 + logO/H.
The galaxies I~Zw~18 and SBS~0335$-$052E are encircled.
}
\label{fig8}
\end{figure}
%%%%%%%%%%%%%%%%%%%%%%%%%%%%%%%%%%%%%%%%%%%%%%%%%%%%%%%%%%%%%%%%

\subsubsection{Dust temperatures}

In Fig. \ref{fig8}, relations among the oxygen abundances
12~+~logO/H and the temperatures of cold, and warm dust
for galaxies from the Herschel sample are shown. Temperatures
vary in the range of 25 K $< T_{\rm cd} <$ 57 K with the average value of 
$\sim$ 30K 
for the cold component and in the range of 72 K $< T_{\rm wd} <$ 129 K with the 
average value of $\sim$ 90K for the warm component. No obvious trends with 
metallicity are found, excluding the two most deviant
and most metal-deficient BCDs, I~Zw~18 and SBS~0335$-$052E.
This result confirms
that found for the temperatures of warm dust in SDSS sample galaxies
(Fig. \ref{fig1}). We note, however, that the average temperatures for 
the SDSS sample are higher, $\sim$120K for the warm component. 
The cause for this discrepancy
remains unclear. However, the temperatures derived for the warm
component in the Herschel sample are more robust,
because SED fitting in these objects is controlled by the photometric
data from WISE, Spitzer, and Herschel and by 
Spitzer/{\it IRS} spectra, while only WISE 
photometric data are available for the SDSS sample. 
We also cannot exclude imperfect correction
of WISE fluxes for the contribution of the nebular emission
lines, which was done with the use of Cloudy photoionised H~{\sc ii} region
models. 
%Overall, we conclude that the SED fits obtained with optical and IR 
%spectra are more reliable compared to that with only photometric data.

\subsubsection{Dust luminosities}

%%%%%%%%%%%%%%%%%%%%%%%%%%%%%%%%%%%%%%%%%%%%%%%%%%%%%%%%%%%%%%%%
% Fig 9 dust luminosities vs L(Hb)
%%%%%%%%%%%%%%%%%%%%%%%%%%%%%%%%%%%%%%%%%%%%%%%%%%%%%%%%%%%%%%%%
\begin{figure}
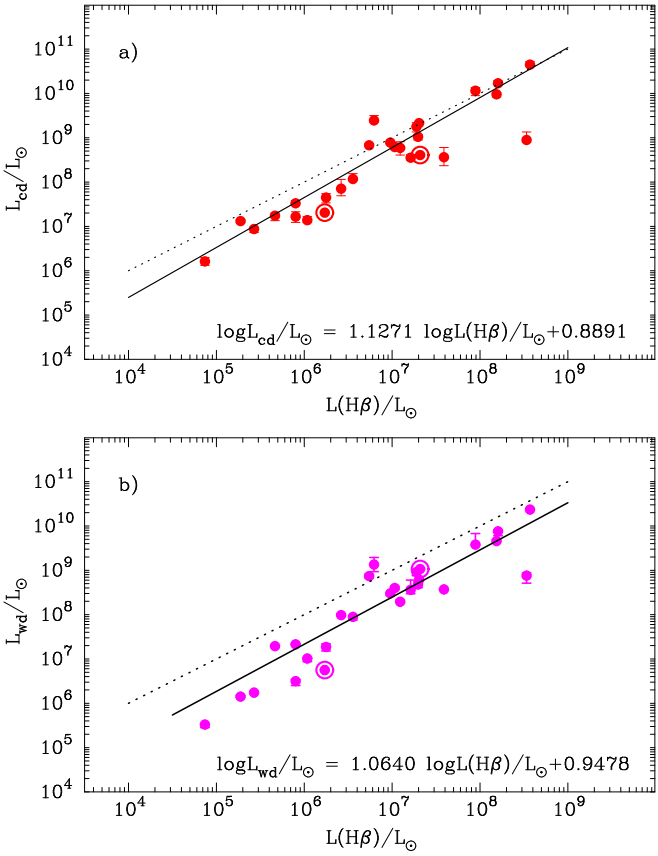

\hbox{
\includegraphics[width=5.5cm,angle=-90]{llow_lhb.ps}
}
\vspace{0.2cm}
\hbox{
\includegraphics[width=5.5cm,angle=-90]{lw_lhb.ps}
}
%\vspace{0.2cm}
%\hbox{
%\includegraphics[width=5.5cm,angle=-90]{lh_lhb.ps}
%}
%\hspace*{0.0cm}\psfig{figure=llow_lhb.ps,angle=-90,width=8.8cm,clip=}
%\vspace{0.1cm}
%\hspace*{0.0cm}\psfig{figure=lw_lhb.ps,angle=-90,width=8.8cm,clip=}
%\vspace{0.1cm}
%\hspace*{0.0cm}\psfig{figure=lh_lhb.ps,angle=-90,width=8.8cm,clip=}
%}
\caption{Dependencies of (a) cold and (b) warm dust luminosities 
on H$\beta$ luminosities for galaxies from the Herschel sample. 
Solid lines are linear maximum-likelihood relations,
while dotted lines are one-to-one relations between dust luminosities and
total luminosities of ionising radiation $L_{\rm ion}$ $\sim$ 100$L$(H$\beta$).
I~Zw~18 and SBS~0335$-$052E are encircled.}
\label{fig9}
\end{figure}
%%%%%%%%%%%%%%%%%%%%%%%%%%%%%%%%%%%%%%%%%%%%%%%%%%%%%%%%%%%%%%%%

Dependencies of cold and warm dust luminosities on the H$\beta$
luminosity $L$(H$\beta$) are shown in Fig. \ref{fig9}. The relation
in Fig. \ref{fig9}a is almost as tight as that in Fig. \ref{fig9}b for
only compact galaxies with Spitzer/{\it IRS} spectra.
This suggests that likely the same sources 
heat the warm and cold dust in compact galaxies from the Herschel sample.
%, indicating that most 
%of the detected dust is located inside the H~{\sc ii} regions, not in the 
%neutral gas.

\subsubsection{Dust masses}

%%%%%%%%%%%%%%%%%%%%%%%%%%%%%%%%%%%%%%%%%%%%%%%%%%%%%%%%%%%%%%%%
% Fig 10 dust-to-stellar mass vs 12+logO/H
%%%%%%%%%%%%%%%%%%%%%%%%%%%%%%%%%%%%%%%%%%%%%%%%%%%%%%%%%%%%%%%%
\begin{figure}
\hbox{
\includegraphics[width=5.5cm,angle=-90]{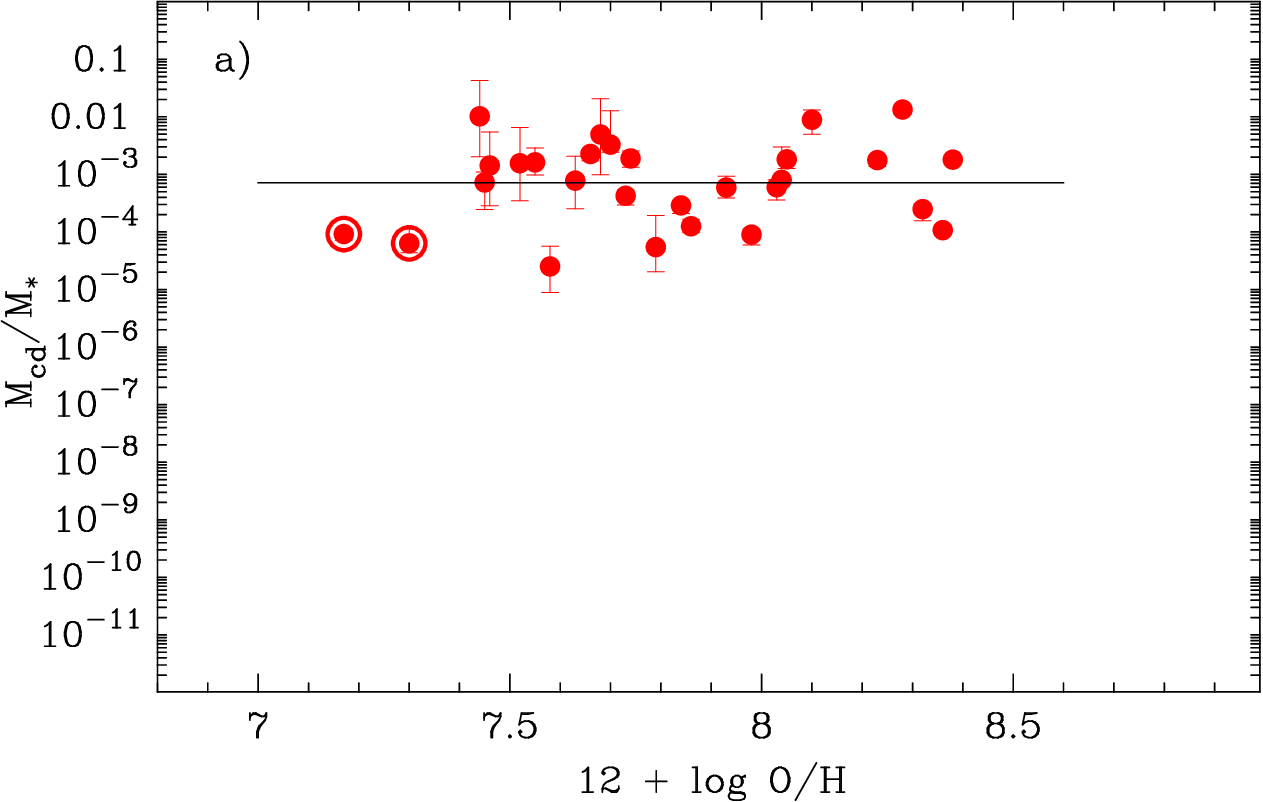}
}
\vspace{0.2cm}
\hbox{
\includegraphics[width=5.5cm,angle=-90]{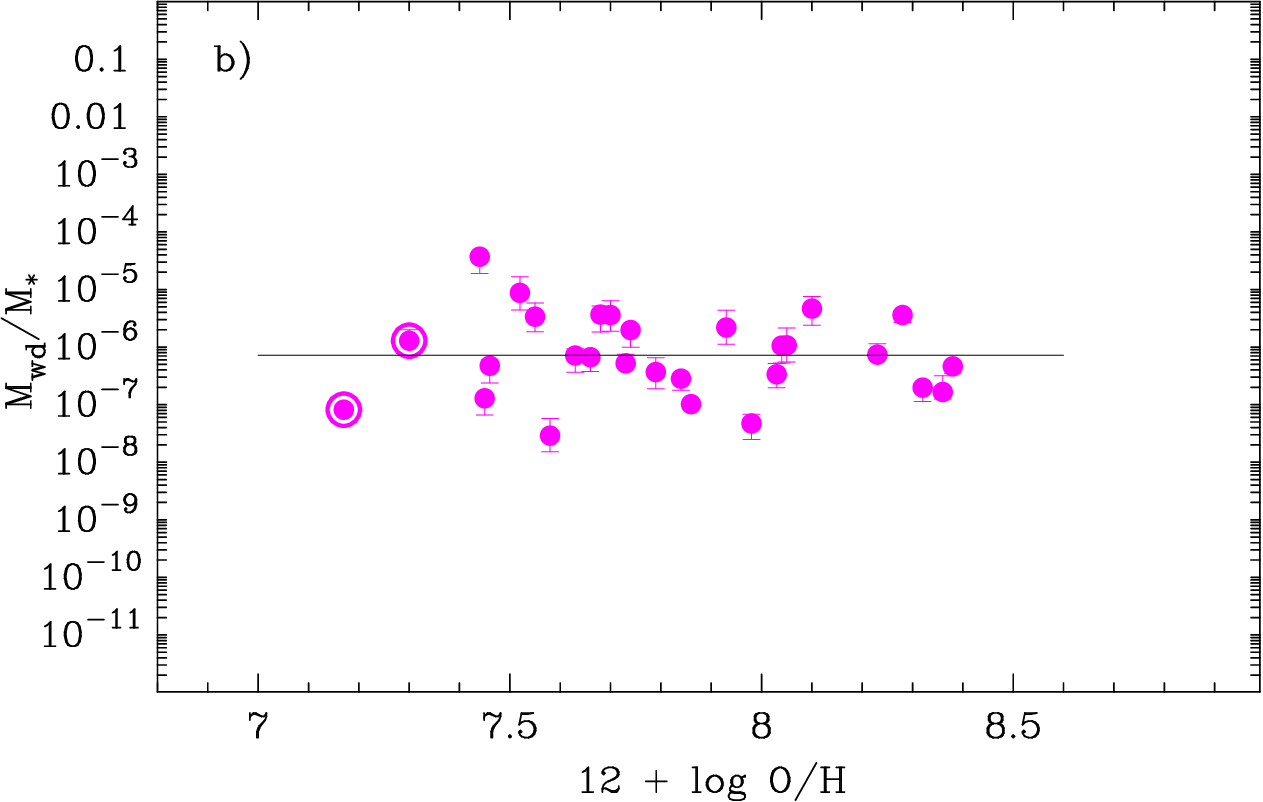}
}
%\vspace{0.2cm}
%\hbox{
%\includegraphics[width=5.5cm,angle=-90]{mh_mtot_oh.ps}
%}
%\hspace*{0.0cm}\psfig{figure=ml_mtot_oh.ps,angle=-90,width=8.8cm,clip=}
%\vspace{0.1cm}
%\hspace*{0.0cm}\psfig{figure=mw_mtot_oh.ps,angle=-90,width=8.8cm,clip=}
%\vspace{0.1cm}
%\hspace*{0.0cm}\psfig{figure=mh_mtot_oh.ps,angle=-90,width=8.8cm,clip=}
%}
\caption{Dependencies of the dust mass-to-total stellar mass ratio on
oxygen abundance are shown for (a) cold and (b) warm dust in galaxies 
from the Herschel sample. The black solid lines in (a) and (b)
indicate average values of $M_{\rm cd}$/$M_{\rm *}$ and $M_{\rm wd}$/$M_{\rm *}$,
respectively.
I~Zw~18 and SBS~0335$-$052E are encircled.
}
\label{fig10}
\end{figure}
%%%%%%%%%%%%%%%%%%%%%%%%%%%%%%%%%%%%%%%%%%%%%%%%%%%%%%%%%%%%%%%%

%%%%%%%%%%%%%%%%%%%%%%%%%%%%%%%%%%%%%%%%%%%%%%%%%%%%%%%%%%%%%%%%
% Fig 11 dust-to-young stellar population mass vs 12+logO/H
%%%%%%%%%%%%%%%%%%%%%%%%%%%%%%%%%%%%%%%%%%%%%%%%%%%%%%%%%%%%%%%%
\begin{figure}
\hbox{
\includegraphics[width=5.5cm,angle=-90]{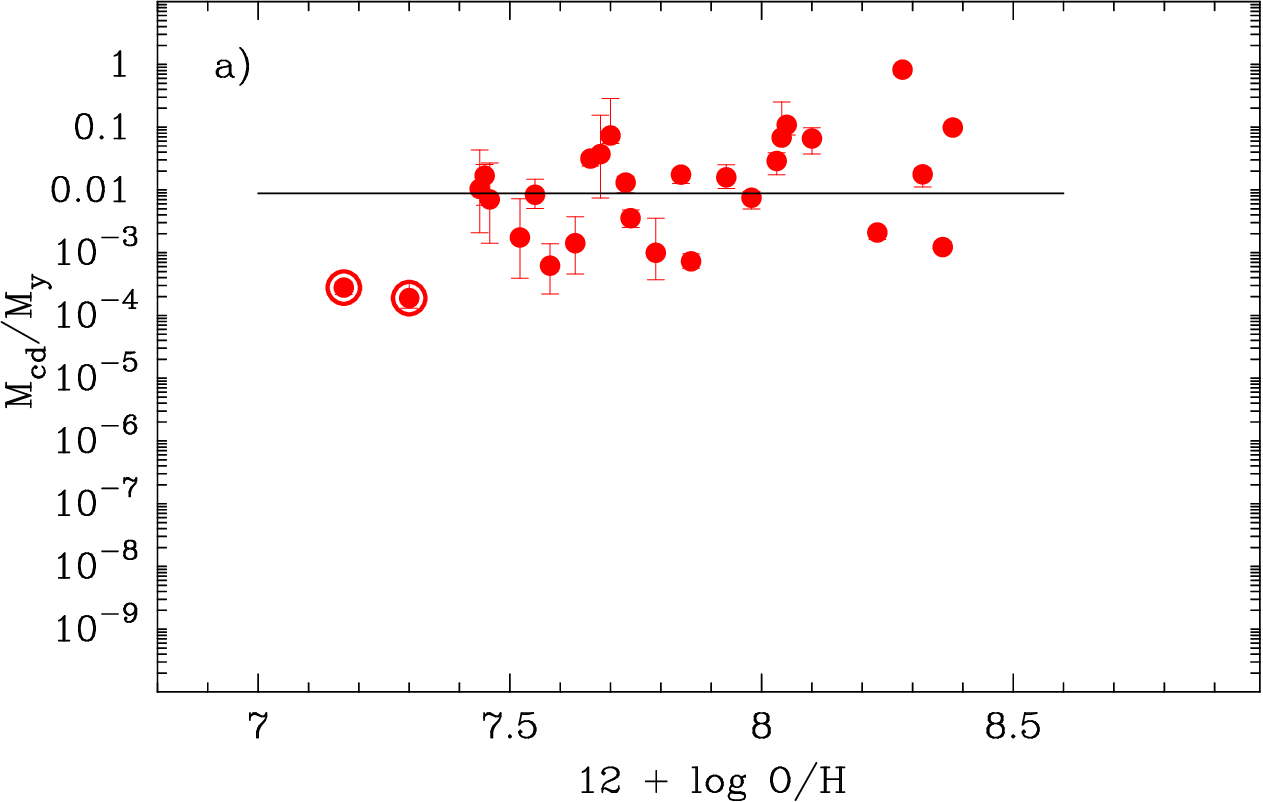}
}
\vspace{0.2cm}
\hbox{
\includegraphics[width=5.5cm,angle=-90]{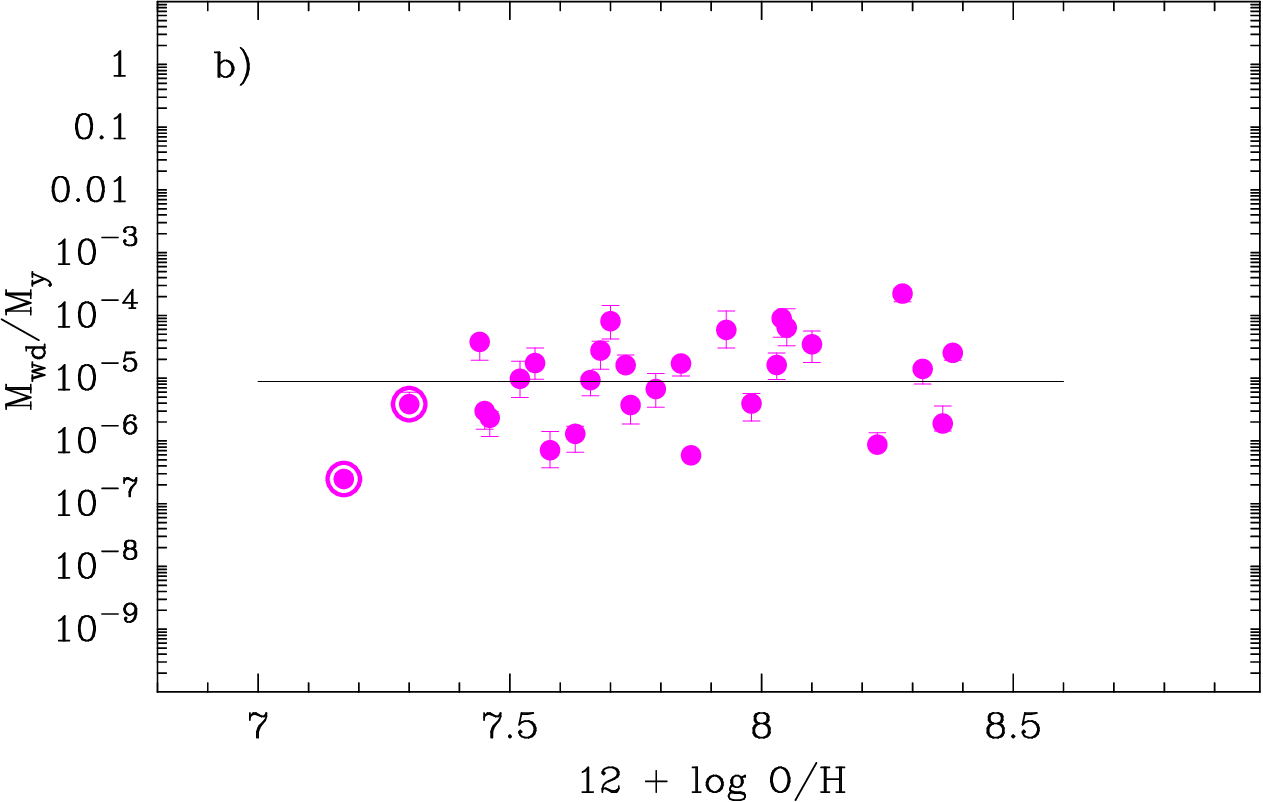}
}
%\vspace{0.2cm}
%\hbox{
%\includegraphics[width=5.5cm,angle=-90]{mh_myoung_oh.ps}
%}
%\hspace*{0.0cm}\psfig{figure=ml_myoung_oh.ps,angle=-90,width=8.8cm,clip=}
%\vspace{0.1cm}
%\hspace*{0.0cm}\psfig{figure=mw_myoung_oh.ps,angle=-90,width=8.8cm,clip=}
%\vspace{0.1cm}
%\hspace*{0.0cm}\psfig{figure=mh_myoung_oh.ps,angle=-90,width=8.8cm,clip=}
%}
\caption{Dependencies of the dust mass-to-young stellar mass ratio on
oxygen abundance are shown for (a) cold and (b) warm dust in galaxies 
from the Herschel sample. The black solid lines in (a) and (b)
indicate average values of $M_{\rm cd}$/$M_{\rm y}$ and $M_{\rm wd}$/$M_{\rm y}$,
respectively.
I~Zw~18 and SBS~0335$-$052E are encircled.
}
\label{fig11}
\end{figure}
%%%%%%%%%%%%%%%%%%%%%%%%%%%%%%%%%%%%%%%%%%%%%%%%%%%%%%%%%%%%%%%%

In Fig. \ref{fig10}, dust mass to total stellar mass ratios on
oxygen abundance 12 + logO/H for cold and warm dust components are shown.
No clear dependence is present. The two most-metal deficient galaxies, 
I~Zw~18 and SBS~0335$-$052E are among the galaxies 
of the Herschel sample with the lowest cold dust mass
(Table \ref{tab1}). In Fig. \ref{fig10}a, they have
low $M_{\rm cd}$/$M_*$ mass ratios of $\sim$10$^{-4}$, which are lower by a 
factor of $\sim$10
than the average value for the sample. This, however, does not deviate 
more than by 1$\sigma$ from other sample galaxies. 
%Deviations on $M_ {\rm wd}$/$M_*$ vs
%12+logO/H diagram (Fig. \ref{fig10}b) of I~Zw~18 and SBS~0335$-$052E
%on $M_ {\rm hd}$/$M_*$ vs 12+logO/H diagram (Fig. \ref{fig10}c) of SBS~0335$-$052E
%are much smaller.

Deviations of I~Zw~18 and SBS~0335$-$052E are much larger in the
$M_ {\rm cd}$/$M_{\rm y}$ vs 12+logO/H diagram (Fig. \ref{fig11}a). 
This is presumably because of the low cold dust mass and high fraction of 
$\sim$30\% of the young stellar population in the total mass, indicating that
both galaxies are relatively unevolved and had no time to
produce dust in large quantities. 
Although the old stellar population with age $\ga$1 Gyr is present 
in I~Zw~18, its mass of $\ga$2$\times$10$^6$ $M_\odot$ \citep{A13} is small
compared to the H~{\sc i} mass of 10$^8$ $M_\odot$ \citep{L12}.
There is no direct evidence for an old stellar population in SBS 0335$-$052E.
\citet{P04} inferred from the integrated light of SBS 0335$-$052E that the 
stellar-to-total baryon mass fraction in this galaxy is 0.035.

Therefore, $M_ {\rm cd}$/$M_{\rm y}$ in these galaxies
is only by $\sim$ 3 times higher than $M_ {\rm cd}$/$M_*$ (Table \ref{tab1}). 
On the other hand, the young-to-total stellar mass ratio is much smaller 
in many other galaxies from the Herschel sample, elevating 
their dust-to-young stellar mass ratio, 
as compared to that in Fig. \ref{fig10}a. The difference 
between I~Zw~18 and SBS~0335$-$052E and other galaxies is lower for the warm
dust component.

%%%%%%%%%%%%%%%%%%%%%%%%%%%%%%%%%%%%%%%%%%%%%%%%%%%%%%%%%%%%%%%%
% Fig 12 dust-to-dust mass relations
%%%%%%%%%%%%%%%%%%%%%%%%%%%%%%%%%%%%%%%%%%%%%%%%%%%%%%%%%%%%%%%%
\begin{figure}
\hbox{
\includegraphics[width=5.5cm,angle=-90]{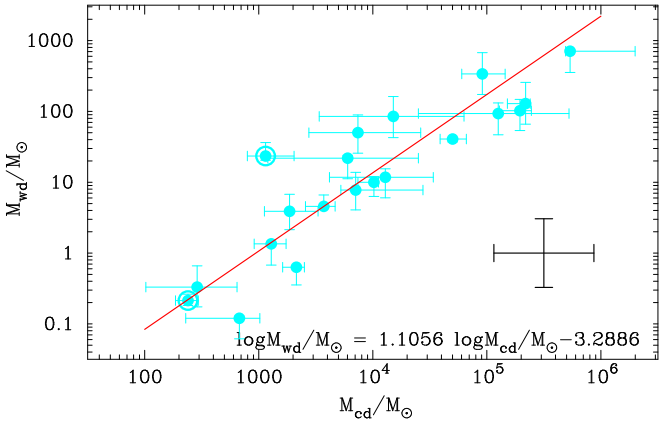}
}
%\vspace{0.2cm}
%\hbox{
%\includegraphics[width=5.5cm,angle=-90]{mh_ml.ps}
%}
%\vspace{0.2cm}
%\hbox{
%\includegraphics[width=5.5cm,angle=-90]{mh_mw.ps}
%}
%\hspace*{0.0cm}\psfig{figure=ml_mw_1.ps,angle=-90,width=8.8cm,clip=}
%\vspace{0.1cm}
%\hspace*{0.0cm}\psfig{figure=mh_ml.ps,angle=-90,width=8.8cm,clip=}
%\vspace{0.1cm}
%\hspace*{0.0cm}\psfig{figure=mh_mw.ps,angle=-90,width=8.8cm,clip=}
%}
\caption{Mass of warm dust vs. mass of cold dust.
The red solid line represents the maximum-likelihood linear regression.
I~Zw~18 and SBS~0335$-$052E are encircled.
}
\label{fig12}
\end{figure}
%%%%%%%%%%%%%%%%%%%%%%%%%%%%%%%%%%%%%%%%%%%%%%%%%%%%%%%%%%%%%%%%

In Fig. \ref{fig12}, the relation is shown between the warm and cold dust masses.
%Note that Spitzer spectra allow a more robust fit of dust 
%emission and in particular to detect hot dust emission, which may be lost
%in photometric data.
%Relations between different masses have similar slopes,
%$M_{\rm wd}$ $\propto$ $M_{\rm cd}$$^{1.19}$,
%$M_{\rm hd}$ $\propto$ $M_{\rm cd}$$^{1.17}$, and 
%$M_{\rm hd}$ $\propto$ $M_{\rm wd}$$^{1.17}$. However, the relation between
%$M_{\rm hd}$ and $M_{\rm cd}$ (Fig. \ref{fig12}b) is not as tight as other two
%relations. 
The presence of a correlation between cold and warm dust
masses for the Herschel sample suggests that these components
are linked to young stellar populations. 
This correlation with dispersion of $\sim$ 0.5 dex opens the opportunity
for estimating the cold dust mass from the warm dust mass. 
The most outlying galaxy in Fig. \ref{fig12} is SBS~0335$-$052E
with highest $M_ {\rm wd}$/$M_{\rm cd}$ ratio. Given the warm dust mass and 
using the relation in Fig. \ref{fig12} would overestimate the cold dust mass by
one order of magnitude for this galaxy. On the other hand,
the relation in the Figure nicely reproduces the cold dust mass in I~Zw~18.

Only a small
fraction of the local star-forming galaxies with low metallicity was observed 
with Herschel \citep{RR13}. On the other hand, there are many more 
compact star-forming galaxies that were detected with WISE
\citep[this paper, ][]{I11b,I14}. The relation
in Fig. \ref{fig12} allows us to estimate the cold dust mass 
in these galaxies with an
uncertainty better than by a factor of three.

\subsubsection{Dust-to-gas mass ratio}

The dust-to-gas mass ratio is often considered as a measure of
metallicity. In this respect it might be compared to the
oxygen abundance 12 + logO/H measured in the ionised gas. One would expect that
the dust-to-gas mass ratio and heavy element mass fraction $Z_{\rm ion}$ in the
ionised gas should be comparable if the galaxy ISM is well mixed.

However, there is some evidence that the neutral gas metallicity is 
much lower than that of the ionised gas. \citet{T02,T05}, \citet{L04,L13},
\citet{LE04} using {\it FUSE} and the Cosmic Origin Spectrograph (COS) onboard 
{\it HST} showed that the oxygen-to-hydrogen abundance ratio in the neutral gas
of low-metallicity BCDs is on average several times lower than that in 
H~{\sc ii} regions. It was even suggested in some papers that
some fraction of pristine neutral gas might be present in these galaxies.

Based on IRAS observations, \citet{LF98} found that the dust-to-gas mass
ratio in dwarf irregulars and BCDs is steeply decreased with decreasing
ionised gas metallicity. Later, \citet{E08} using Spitzer observations
confirmed this result. 
A detailed study of the dust-to-gas mass ratio in the galaxies from
the Herschel sample was done by \citet{RR14}. Based on the galaxies
from \citet{RR13}, they compiled H~{\sc i} data and calculated 
the mass of the molecular hydrogen by using the CO-to-H$_2$ conversion factor
for galaxies with detected CO emission and by extrapolating the conversion
factor for galaxies with low metallicity, where CO emission was not
detected. \citet{RR14} found that the dust-to-gas mass ratio is steeply 
decreased with decreasing metallicity, a result 
similar to that obtained by \citet{LF98} and \citet{E08}.
%We wish to check whether the same is valid 
%for the Herschel sample galaxies.

\begin{figure*}
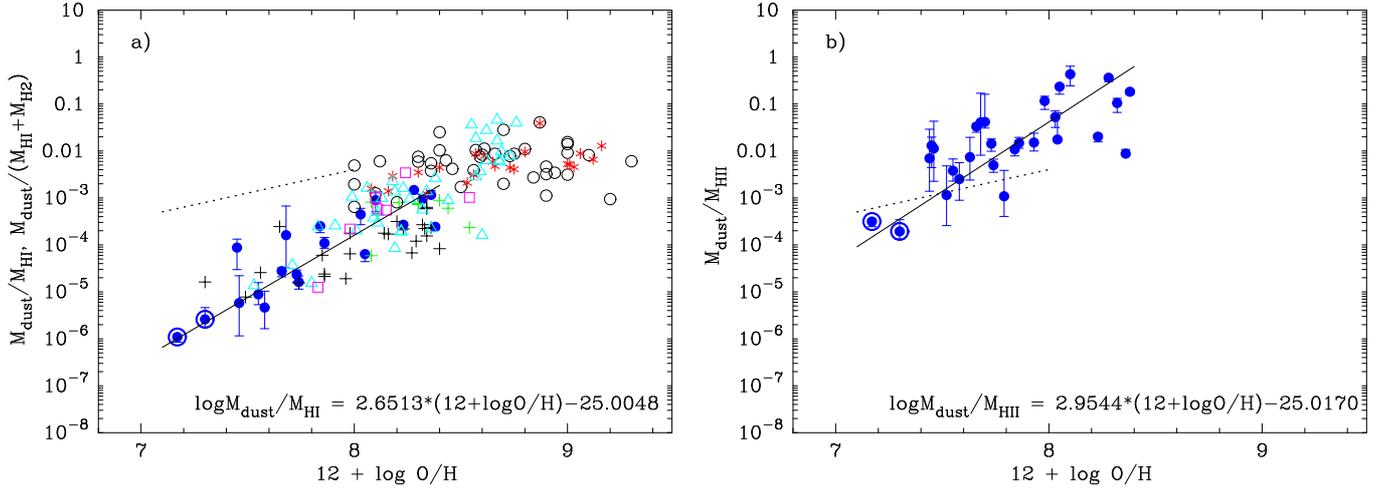

\includegraphics[width=6.4cm,angle=-90]{mh21_mdust_2.ps}
\hspace{0.3cm}\includegraphics[width=6.4cm,angle=-90]{mion_mdust_2.ps}
\caption{(a) Dependence of the dust-to-hydrogen mass ratio relative
to the oxygen abundance 12 + logO/H. The data from this paper are shown by
large filled blue circles and are fitted by the linear maximum-likelihood
regression (black solid line). The two lowest-metallicity galaxies,
I~Zw~18 and SBS~0335--052E, are encircled. 
%The black dotted line denotes 
%the one-to-one
%line between dust-to-gas mass ratio and the heavy-element mass fraction 
%$Z_{\rm ion}$ (see Eq. \ref{Z}) 
%in the ionised gas as derived from the oxygen abundance.
The dust-to-gas ratio from a compiled sample of dwarf irregulars and blue
compact dwarf (BCD) galaxies by \citet{LF98} (IRAS data) are denoted by 
black crosses 
\citep[note that for GR8 we adopt the oxygen abundance from ][]{M11}. 
The data for 16 BCDs (IRAS data) by \citet{H02} are shown by magenta
open squares. The dust-to-gas ratio
for 66 star-forming galaxies observed with Spitzer 
by \citet{E08} are shown by light blue triangles.
The data compiled from literature for a wide sample of galaxies using different
 submm ($>$ 160$\mu$m) observations \citep{Ga11} 
 are denoted by large open black circles. The available data 
gathered by \citet{J02} and their observations with \textit{SCUBA} are 
shown by red asterisks.
(b) Dependence of the dust-to-ionised-hydrogen mass ratio relative
to the oxygen abundance 12 + logO/H. The data from this paper are shown by
large filled blue circles and are fitted by the linear maximum-likelihood
regression (black solid line). The two lowest-metallicity galaxies,
I~Zw~18 and SBS~0335--052E, are encircled. The black dotted lines 
in both panels denote a one-to-one correspondence
between the dust-to-gas mass ratio and the heavy-element mass fraction 
$Z_{\rm ion}$ (see Eq. \ref{Z}) 
in the ionised gas as derived from the oxygen abundance.
 }
\label{fig13}
\end{figure*}

In Fig. \ref{fig13}a, we show the dust-to-gas mass - oxygen abundance 
diagram for the Herschel
sample of compact galaxies together with data from 
the literature, which include both dwarf and giant galaxies. 
The mass of molecular hydrogen $M_{\rm H2}$ for three galaxies from
the Herschel sample with detected CO emission (Haro 11, Mrk 1089, Mrk 930)
was taken from \citet{Co14} and added to the H~{\sc i} mass.
The $M_{\rm H2}$/$M_{\rm HI}$ ratios for Mrk 1089 and Mrk 930 are low, of 
$\sim$ 0.2 -- 0.3. However, this ratio of $\sim$5 is much higher for Haro 11.
For galaxies selected from the literature, 
we show the data in Fig. \ref{fig13}a,
which include the H$_2$ mass, whenever it was available.
We do not use the extrapolation of the CO-to-H$_2$ conversion factor to
low metallicities to derive the H$_2$ mass in 
galaxies from the Herschel sample with undetected
CO emission because this procedure is highly uncertain.

We confirm a sharp decrease of dust-to-gas mass ratios
with decreasing metallicity at 12 + logO/H $<$ 8.5, while the ratio is almost 
constant in giant galaxies with higher metallicities. The Herschel
data can be fit by a linear maximum-likelihood relation
shown as a black solid line. We note the steep slope of $\sim$ 2.65, which
is steeper than the linear extrapolation of the data for higher-metallicity
galaxies by \citet{D07} and the model predictions by \citet{H02}.
%who considered dust destruction by the SNe remnants.
We note that the dust-to-gas mass ratios in Fig. \ref{fig13}a are global 
characteristics of galaxies because they are derived from the total dust and
gas masses.

The dotted line in Fig. \ref{fig13}a is the one-to-one line between
the dust-to-gas mass ratio and the heavy-element mass fraction 
$Z_{\rm ion}$ in
the ionised gas. $Z_{\rm ion}$ is derived from the relation
\begin{equation}
Z_{\rm ion}=16\times\left(\frac{\rm O}{\rm H}\right)\left(\frac{Z}{Z_{\rm O}}\right), \label{Z}
\end{equation}
where $Z$/$Z_{\rm O}$ is the ratio by mass of all heavy elements to oxygen,
which we adopt to be 2.5. O/H is the oxygen abundance by number derived
from the H~{\sc ii} region's optical spectrum. 

This relation shows that the observed
dust-to-gas mass ratio in dwarf galaxies is by a factor of
$\sim$ 100 lower than the heavy element mass fraction in the dwarf's
H~{\sc ii} regions.
Keeping in mind that the metallicity of the neutral gas in these galaxies
can be lower by a factor of up to $\sim$10 than the heavy element mass 
fraction in H~{\sc ii} regions \citep[e.g. ][]{T05}, 
we conclude that the global dust-to-gas mass ratio in dwarf galaxies
% \textbf{i.e. metallicity of dwarf galaxies as whole} 
is $\sim$ 10 times lower than the neutral gas
metallicity. From the two relations (Fig. \ref{fig13}a), 
we can derive the dust-to-metal mass ratio, which 
declines with decreasing metallicity. This is in line with the trends in 
%dust-metallicity ratios
dust-to-metal ratios (i.e. the mass fraction of metals in dust) found 
by \citet{D13} for low-metallicity gamma-ray burst (GRB) hosts and 
damped Ly-$\alpha$ absorbers (DLAs).

The lowest global dust-to-gas mass ratios are derived for the two 
lowest-metallicity
BCDs, I~Zw~18 and SBS~0335$-$052E, shown in Fig. \ref{fig13}a.
They are in nice agreement with dust-to-gas mass ratios 
obtained for these galaxies by \citet{RR14} and 
for some other most-metal poor galaxies with 12 + logO/H $\la$ 7.6
\citep[this paper, ][]{LF98,E08}. \citet{H13} derived a similar 
dust-to-gas mass ratio for I~Zw~18, but at a value $\sim$5000 times higher for
SBS~0335$-$052E. The difference in the case of SBS~0335$-$052E is in part due 
to that \citet{H13} derived a dust mass, which is a factor of $\sim$40 higher
than the value in this paper (Table \ref{tab1}), the latter being consistent
with \citet{RR13}. Furthermore, \citet{H13} assumed that the dust is located
inside an $\sim$1\arcsec\ aperture, which is
much smaller than the larger aperture, of $\sim$10\arcsec\ in radius, where
Herschel fluxes were measured. We adopt a large aperture 
of $\sim$10\arcsec. \citet{H13} also used 
the H~{\sc i} mass  inside $\sim$1\arcsec\ aperture, 
while we use the total H~{\sc i} mass of
4.2$\times$10$^8$ $M_\odot$ for SBS~0335$-$052E \citep{E09}. 
     Furthermore, hydrogen in the central part of the 
galaxy within a 1\arcsec\ aperture is mostly ionised, and its mass that adopts 
the electron number density of $\sim$100 cm$^{-3}$ is by at least a factor of 
$\sim$10 larger than the mass of neutral gas within the same aperture. 
%Furthermore, dust can be localised not only in the neutral gas
%region, but also in the ionised gas region. 
In these circumstances, ionised
gas should be taken into account in the dust-to-gas mass ratio determination.
On the other hand, the global dust-to-gas mass ratio is almost insensitive to
the presence of ionised gas, because its total mass is much lower that the
neutral gas mass.
 
In this respect,
our dust-to-gas mass ratio for SBS~0335$-$052E can be considered as a global
characteristic of the galaxy, while that of \citet{H13} 
describes a local characteristic for the central part of the galaxy.

In Fig. \ref{fig13}b, we show the dust-to-ionised-gas mass ratio for
our galaxies. The ionised gas mass is derived using relation
\begin{equation}
M({\rm H\ II}) = \frac{L({\rm H}\beta)m_p}{\epsilon N_{\rm e}},
\end{equation}
where $m_p$ is the proton mass, $N_{\rm e}$ is the electron number density, 
$L$(H$\beta$) is the extinction-corrected total H$\beta$ luminosity, and
$\epsilon$ is the H$\beta$ emissivity, as defined by equation \citep{A84}
\begin{equation}
\epsilon = 1.37\times 10^{-25} t_{\rm e}^{-0.982}\exp(-0.104/t_{\rm e}),
\end{equation}
$t_{\rm e}$ = 10$^{-4}$$T_{\rm e}$.

It is seen in Fig. \ref{fig13}b that the dust-to-ionised-gas mass ratio is
sharply increased with metallicity and generally is higher than the 
dotted line denoting a ratio of unity 
between $M_{\rm dust}$/$M_{\rm H\ II}$ and $Z_{\rm ion}$. 
Only in the lowest-metallicity galaxies are they 
comparable. This suggests that although radiation by a young 
stellar population is the main source of dust heating, most of dust is
located in the neutral gas, and the fraction of dust in the neutral gas 
is increased with increasing metallicity.

%%%%%%%%%%%%%%%%%%%%%%%%%%%%%%%%%%%%%%%%%%%%%%%%%%%%%%%%%%%%%%%%
% Fig 13 dust-to-hydrogen mass relations
%%%%%%%%%%%%%%%%%%%%%%%%%%%%%%%%%%%%%%%%%%%%%%%%%%%%%%%%%%%%%%%%
%%%%%%%%%%%%%%%%%%%%%%%%%%%%%%%%%%%%%%%%%%%%%%%%%%%%%%%%%%%%%%%%

\subsubsection{Sub-mm excess in star-forming dwarf galaxies}

\citet{RR13} extensively discussed the sub-mm excess at wavelengths
$\ga$500 $\mu$m, which is detected in a significant fraction of galaxies from 
the Herschel sample. They noted that this excess is more frequent
in low-metallicity galaxies. In particular, they discuss two galaxies with
this excess, Haro 11 and II Zw 40. 
\citet{RR13} also detected the sub-mm excess in the galaxies 
HS 0052$+$2536 and VII Zw 403 from our Herschel sample.
Furthermore, \citet{RR13} analyse different
mechanisms, which may cause this emission. All of them are related to
dust emission.

On the other hand, the presence of cold dust in 
low-metallicity compact dwarf galaxies
with active star formation and intense
UV stellar radiation is difficult to understand. We suggest that the 
free-free emission may play a role
in producing sub-mm excess in these galaxies with strong nebular continuum
in the optical range and at longer wavelengths, as is indicated by
a high equivalent width of the H$\beta$ emission line. It was noted in 
Sect. \ref{SED} that the fluxes of the free-free emission in sub-mm and radio
ranges were calculated from the H$\beta$ emission line flux, according
to \citet{CD86}. We compare predicted and observed 
fluxes in the cm range where emission is dominated by free-free processes.
For most objects from our Herschel sample, no data are available in
this wavelength range. For the remaining galaxies, Haro 11 (Fig. \ref{fig6}-a1),
II Zw 40 (Fig. \ref{fig6}-b1), I Zw 18 (Fig. \ref{fig6}-b2), 
Mrk 1089 (Fig. \ref{fig6}-b3), Mrk 209 (Fig. \ref{fig6}-b4),  
Mrk 930 (Fig. \ref{fig6}-b7), and
SBS 0335$-$052E (Fig. \ref{fig6}-c2),
the agreement is good.

Depending on the galaxy properties, the contribution of free-free emission 
may be significant in the sub-mm or mm wavelength ranges (Fig. \ref{fig6}).
We note that the role of free-free emission in producing sub-mm excess
was discussed earlier by \citet{Ga03} and 
\citet{G05} for four high-metallicity star-forming galaxies, NGC 1569, 
II Zw 40, He 2-10, and NGC 1140. They concluded that free-free emission is 
insufficient to explain this excess and that emission of very cold dust
is needed. However, we point out that \citet{RR13} did not detect the excess in 
NGC~1140, while it is present with the \citet{G05} data. 
We also conclude from our SED fitting (Fig. \ref{fig6}-b8) that a 
sub-mm excess in this galaxy is not apparent.

We note that sub-mm and mm observations for II Zw 40 are controversial
(Fig. \ref{fig6}-b1). Submillimetre array (SMA) observations at
880$\mu$m \citep{H11} and MRT observations at 1.25 mm \citep{A04} are in
very nice agreement with our fit, while 850$\mu$m SCUBA and 1.3mm MAMBO
observations by \citet{G05} and \citet{H05} are not. 
It is possible that differences in observed fluxes are due to the
fact that II~Zw~40 is located at low Galactic latitude with a high Milky Way
extinction. This makes subtraction of bright foreground Galactic cirrus
somewhat uncertain. We also note that 
870$\mu$m LABOCA observations by \citet{Ga09} are above the fit for Haro 11
(Figs. \ref{fig6}-a1). On the other hand, observed and predicted 
fluxes at $\lambda$ $\ga$ 850$\mu$m for another two galaxies, 
NGC 1140 (\ref{fig6}-b8) and SBS 0335$-$052E (\ref{fig6}-c2) are
in very good agreement and do not show evidence for a very cold dust.

In Fig. \ref{fig6},
blue vertical dashed lines indicate the wavelengths $\lambda_{\rm ff}$ 
at which contributions
of dust and free-free emission are equal. The most extreme cases are the
most-metal deficient galaxies from the sample, I~Zw~18 and 
SBS~0335$-$052E, with $\lambda_{\rm ff}$ $\sim$ 370$\mu$m and $\sim$ 340$\mu$m,
respectively (Figs. \ref{fig6}-b2 and \ref{fig6}-c2). In particular,
ALMA observations of SBS~0335$-$052E clearly indicate that emission at 
870$\mu$m is almost totally dominated by free-free emission 
\citep[Fig. \ref{fig6}-c2, ][]{H13}. There is also a general tendency of 
lowering
$\lambda_{\rm ff}$ with decreasing oxygen abundance, 12 + logO/H. The 
wavelength $\lambda_{\rm ff}$ for galaxies with 12~+~logO/H $\la$ 8.0 is 
commonly in the sub-mm range, while it is $\ga$ 1mm for higher metallicities
(Fig. \ref{fig6}). This agrees with statistics of the sub-mm excess 
discussed by \citet{RR13}.

Regarding Haro 11 and II Zw 40, $\lambda_{\rm ff}$ are $\sim$ 700$\mu$m and 
$\sim$ 960$\mu$m, respectively (Figs. \ref{fig6}-a1 and \ref{fig6}-b1). 
Therefore, a sub-mm excess in Haro 11 
due to the free-free emission is expected at shorter wavelengths, and it 
is seen by Herschel at 500$\mu$m, while it is not in II Zw 40.

\begin{figure}
\begin{center}
\includegraphics[width=6.5cm,angle=-90]{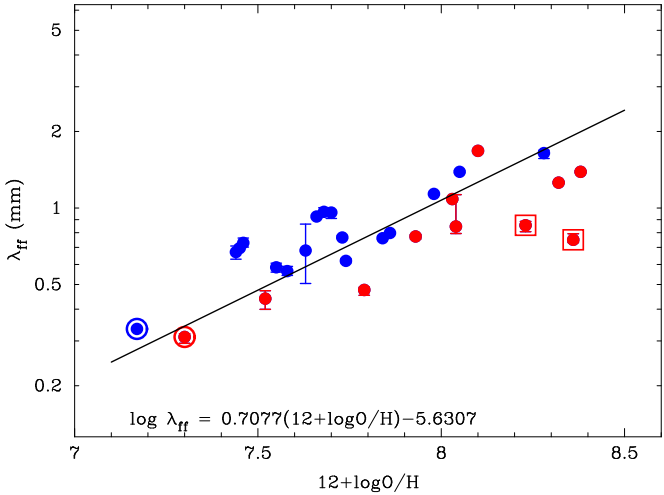}
\end{center}
\caption{\label{fig14} Dependence of the wavelength $\lambda_{\rm ff}$ in 
sub-mm and mm ranges on the oxygen abundance 12+logO/H, at which 
monochromatic dust emission is equal to monochromatic free-free emission.
Red filled circles and blue filled circles are data for galaxies
from the Herschel sample with $L$(H$\beta$) $\geq$ 10$^7$ $L_\odot$ and 
$L$(H$\beta$) $<$ 10$^7$ $L_\odot$, respectively.
% Blue dotted and red
%dashed lines are respective maximum-likelihood linear regressions,
%while 
The black solid line is the maximum-likelihood regression for the
entire Herschel sample. I~Zw~18 and SBS~0335$-$052E are 
encircled, while Haro 11 and II Zw 40 are surrounded by squares.}
\end{figure}

The dependence of $\lambda_{\rm ff}$ on oxygen abundance for the entire
Herschel sample is shown in Fig. \ref{fig14}. A clear increase of
$\lambda_{\rm ff}$ with increasing 12+logO/H is seen, implying that the
sub-mm excess in low-metallicity galaxies 
may be due to the free-free emission from relatively transparent 
H~{\sc ii} regions, which are seen in the optical range.
% No emission of very cold dust is needed for that. 
However, metallicity is not the only parameter
determining $\lambda_{\rm ff}$, which is lower for galaxies with higher
luminosity $L$(H$\beta$) of the H$\beta$ emission line
(compare galaxies shown in Fig. \ref{fig14}). In particular,
$\lambda_{\rm ff}$ for the two most deviant high-metallicity galaxies Haro 11
and II Zw 40 with high $L$(H$\beta$) is also in the sub-mm range.

Unfortunately, there are no observational data in the sub-mm range
for most of the galaxies shown in Fig. \ref{fig6}. Therefore, we cannot check
the importance of free-free emission in producing sub-mm excess in these
low-metallicity galaxies. At least, we can conclude that the wavelength, 
at which the contribution of free-free emission becomes important
is progressively decreased with decreasing metallicity, which is
consistent with measured statistics of the sub-mm excess \citep{RR13}. 
Hopefully, forthcoming observations of these galaxies with ALMA will further
clarify this issue.

%%%%%%%%%%%%%%%%%%%%%%%%%%%%%%%%%%%%%%%%%%%%%%%%%%%%%%%%%%
%  Fig.14 submm excess
%%%%%%%%%%%%%%%%%%%%%%%%%%%%%%%%%%%%%%%%%%%%%%%%%%%%%%%%%
%%%%%%%%%%%%%%%%%%%%%%%%%%%%%%%%%%%%%%%%%%%%%%%%%%%%%%%%%

\subsection{The components of the SBS~0335$-$052E SED in the infrared}

\citet{T99} were the first who noted the existence of highly obscured regions
in SBS~0335$-$052E based on ISO observations. More recently,
\citet{PS02} based on the same observations concluded that extinction
in the obscured region is of order $A_V$ $\sim$ 30 mag and that most of the star
formation in SBS~0335$-$052E is hidden and is not seen even in the 
near-infrared
range. \cite{H04} discovered silicate absorption at 9.7 $\mu$m in the
Spitzer spectrum of SBS~0335$-$052E, which is indicative of the presence
of highly obscured regions. Finally, \citet{H13} modelled the
IR SED using the DUSTY code with inclusion of obscured regions.

%It is to note that the 9.7$\mu$m depression is seen in spectra of some other
%galaxies, but all of them are among the most metal-rich galaxies
%from the Herschel sample. The wavelength of 9.7$\mu$m is indicated in 
%Fig. \ref{fig6} by vertical cyan dashed line.

On the other hand, our fit of free-free emission from the observed H$\beta$
flux in SBS~0335$-$052E is in excellent agreement with the observed fluxes in 
the sub-mm and
radio ranges (Fig. \ref{fig6}-c2). There are many spectroscopic observations of 
SBS~0335$-$052E with H$\beta$ flux measurements of 
$\sim$ (4 -- 10)$\times$10$^{-14}$ erg s$^{-1}$cm$^{-2}$. For the fit 
of free-free emission in Fig. \ref{fig6}-c2, we adopted the value of
1.05$\times$10$^{-13}$ erg s$^{-1}$cm$^{-2}$ by \citet{I06b} for the entire
galaxy. However, if, in SBS~0335$-$052E, most of the star formation is hidden 
by a foreground absorbing layer as suggested by
\citet{PS02} then one would expect higher observed radio 
fluxes, as compared to those predicted from the observed H$\beta$ flux.

Furthermore, \citet{IT11} predicted fluxes of nebular 
[S {\sc iv}] $\lambda$10.51$\mu$m and [Ne {\sc iii}] $\lambda$15.55$\mu$m 
emission lines from optical spectroscopic observations and Cloudy 
photoinised H~{\sc ii} region models, which are in good agreement with fluxes
observed by \citet{H04} in the Spitzer/{\it IRS} spectrum of SBS~0335$-$052E.
Finally, ALMA continuum observations at 870$\mu$m \citep{H13} and
VLA continuum observations in the cm range \citep{J09} do not reveal new
spatially distinct star-forming regions in addition to those seen at shorter 
wavelengths.
 
%\textbf{Note, that the seen depression is in between strong PAH and 
%MIR emission lines. In Fig. \ref{fig6} this 
%%feature 
%depression is seen in 15 galaxies with
%oxygen abundance 12 +log O/H from 8.02 to 8.38 and averaged value 8.22. The 
%mean oxygen abundance for galaxies without 
%%silicate absoption 
%this feature is 7.65 
%(18 galaxies with Spitzer/{\it IRS} spectrum). This is in line with conclusion
%of \citet{D07} that there is threshold in oxygen abundance near 8.1 with lower
%PAH emission for low-metallicity galaxies.}

Therefore, the origin of silicate absorption at 9.7$\mu$m is puzzling as the
observations imply that the contribution of obscured star-forming regions
to FIR, sub-mm, and thermal radio emission should be lower than that 
of those star-forming regions, which are visible in the optical range. 

According to Starburst99 models for instantaneous
bursts \citep{L99}, the H$\beta$ luminosity of visible
star-forming regions in SBS~0335$-$052E, $L$(H$\beta$) = 
1.6$\times$10$^7$ $L_\odot$ corresponds to a luminosity of ionising radiation 
beyond the Lyman limit, $L_{\rm ion}$ $\sim$ 100 $L$(H$\beta$)
= 1.6$\times$10$^9$ $L_\odot$, and a total UV luminosity $L_{\rm UV}$
of 3.9$\times$10$^9$ $L_\odot$. The $V$-band extinction $A$($V$) of 0.3 mag
for SBS~0335$-$052E would correspond to 
$A$(0.1$\mu$m) $\sim$ 3$\times$$A$($V$) $\sim$ 0.9 mag meaning that more than
half of the UV radiation is absorbed by dust. This seems to be sufficient
to explain our derived $L_{\rm FIR}$ of 1.7$\times$10$^9$ $L_\odot$, which
includes cold and warm dust components (Table \ref{tab1}).

\begin{figure}
\begin{center}
\includegraphics[width=6.5cm,angle=-90]{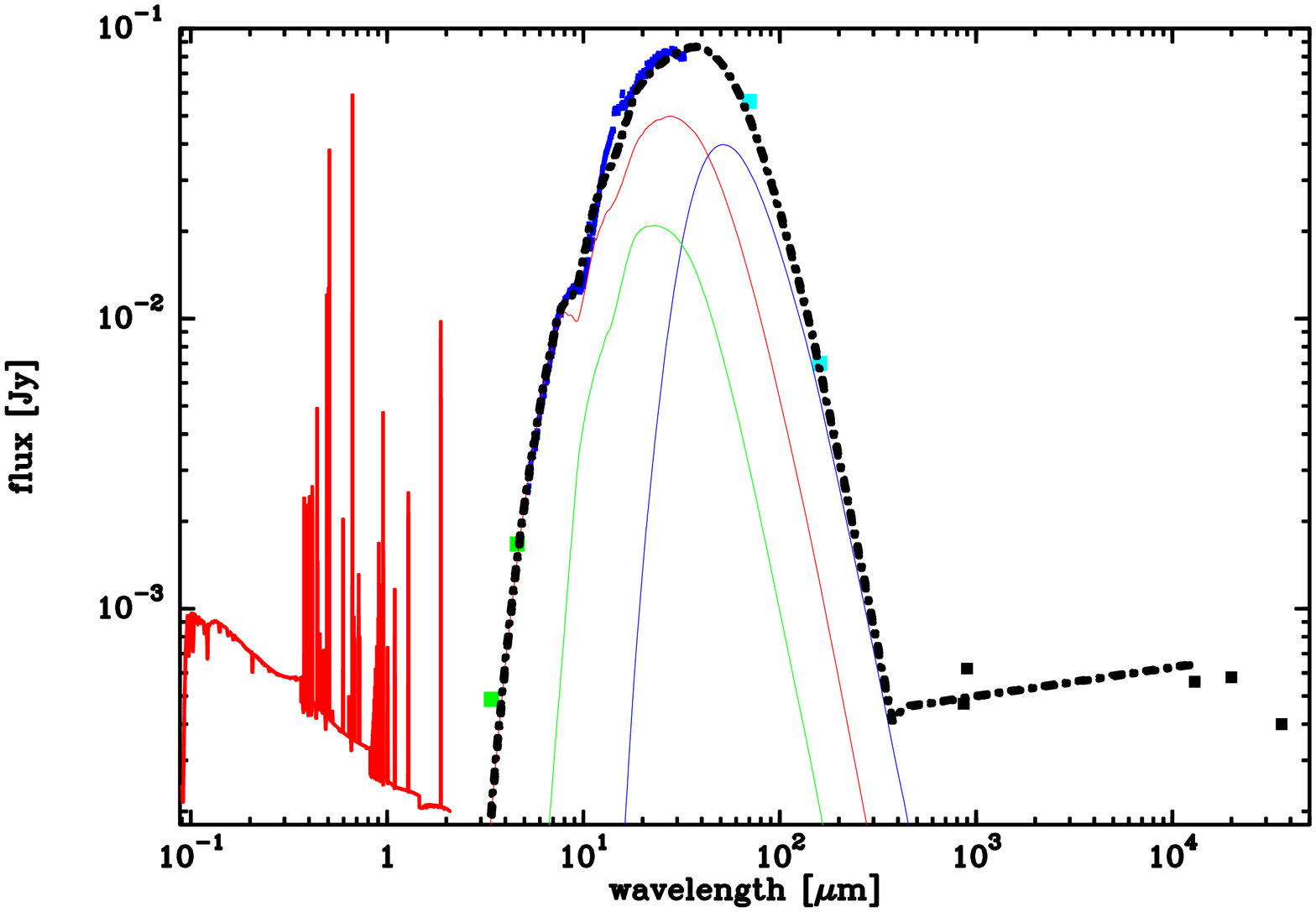}
\end{center}
\caption{\label{fig15} The full spectrum of SBS~0335-052E.  Green
    points: WISE. Blue points: Spitzer. Cyan points:
  Herschel. Black points: submm and radio.  The IR part of the SED
  can be deconvolved into three components: a cold one of $T_1=65$ K
  (blue line), a warm one of $T_2=160$ K (green line), and 
emission from dust surrounding the still
  enshrouded star clusters (red line).  Their sum is shown by dot-dashed
line.}
\end{figure}

Figure\ref{fig15} shows the full spectrum of SBS~0335$-$052E from the Lyman
limit to centimeter radio wavelengths.  Between $3\mu{\rm m} < \lambda
< 0.5$mm, the radiation comes from dust. At shorter wavelengths, it
is stellar and nebular emission including many lines; at longer
wavelengths, it is free-free radiation. Because the radio continuum
becomes opaque ($\tau_{\rm c}=1$) at $\lambda > 3$ cm, this implies an
emission measure EM $\sim 10^9$ pc cm$^{-6}$ or an electron density of
$10^4$ cm$^{-3}$ over a length of 1 pc.

More than half of the total luminosity of the galaxy, 
$L_{\rm gal}=3.9 \times 10^9$
$L_\odot$, is emitted at optical wavelengths
($L_{\rm opt}= 2.2 \times 10^9$ $L_\odot$ for $\lambda < 2\mu$m) and
has not been attenuated by dust. The remaining luminosity, $L_{\rm FIR}=
1.7 \times 10^9$ $L_\odot$, has been processed by dust.  If the galaxy is
powered by O6 stars with having a luminosity $L_* =2.5 \times 10^5$
$L_\odot$, it will take $N_*= 1.56 \times 10^4$ of them to account for
$L_{\rm gal}$.

As one O6 star emits $N_{\rm L} \sim 1.2 \times 10^{49}$ Lyman
continuum photons per s, the output of all O6 stars amounts to $N_*
N_{\rm L}= 1.9 \times 10^{53}$ photons s$^{-1}$. If the fraction
$L_{\rm opt} /L_{\rm gal}$ $\sim$ 22/39 leads to ionisation, one comes up 
with a number of ionising photons $N_{\rm ion} = 1.1 \times 10^{53}$ which
agrees very well with the number derived from the optically thin radio
emission, for example, at $\lambda = 1$ cm \citep{J09}.

By modelling the dust
radiation from a source in a radiative transfer calculation for simple 
geometries, one can sometimes derive
good estimates of the spatial distribution, mass, and temperature range
of the dust.  Here, however, the geometry is obviously much too complex,
and we restrict ourselves to a decomposition of the infrared part of
the SED into components and their semi-quantitative analysis.
Therefore, most of the following numbers (where we use the correct
emissivity $\kappa_\nu B_\nu(T_{\rm dust})$ and not the approximation
$\nu^\beta B_\nu(T_{\rm dust})$) are rough but indicative of the
range of possible values.

In analysing the SED, we start by looking at its coldest component
defined by the far infrared emission.  The data points at 70 and
160$\mu$m imply a colour temperature of about $T_1=65$ K in which the dust
acquires at a distance of $R_1 \sim 100$ pc from a central source of
luminosity $L_{\rm gal}$.  Of course, when $L$ and $R$ are fixed, the
temperature depends on the size and mineralogy of the grain, but $T_1$
is a mean value for the grain's standard size range (0.03 to 0.3$\mu$m) and
standard composition (silicate, amorphous carbon and graphite).  The
mass of this dust component is $M_1 \sim 2000$ $M_\odot$ or 
$\sim$1.5 times higher than the mass derived adopting $\beta$ = 2
(Table \ref{tab1}), and it is located at the edge of or just outside 
the H~{\sc ii} region. If cold dust forms a
shell of width $\Delta R_1$ around the powering star cluster and if
$\rho$ is the density of dust per cm$^3$, then the visual extinction
optical depth through this shell is $\tau_{1, V} = \kappa_{V}\Delta
R_1 \rho = \kappa_{V} M_1/4\pi R_1^2 \sim 0.13$ for an extinction
coefficient at $V$ of $\kappa_{V} \sim 4\times 10^4$ cm$^2$ per g of
dust.  This cool shell is therefore almost transparent to radiation
from inside.

To explain the SED, one needs more warmer components.  An
important guide to their constraint plays the 9.7$\mu$m silicate
feature.  It is claimed to be seen in absorption by \citet{H04} 
with an optical depth $\tau_{9.7\mu{\rm m}} = 0.49$.  One
way to produce it would be by a layer of such an optical depth that
lies in front of a background source with an intrinsic, undiminished
flux of $\sim$0.02 Jy at 9.7$\mu$m.  The foreground layer must be
cooler than 150 K, or it would show the 9.7$\mu$m resonance in
emission. The background source itself, unless it is
optically thick to stellar radiation, should for similar reasons be
cooler than 150 K.

The visual extinction optical depth of the foreground layer is
$\tau_{V} \sim 0.49 \times \tau_{V} / \tau_{9.7\mu{\rm m}}
\sim 7.5$.  If it were in front of the stellar superclusters it would
strongly suppress their optical and UV emission which would then be
intrinsically brighter by a factor $\sim$$e^{7.5}$ than what is observed.
That, however, is excluded because the IR part of the SED accounts for
less than half of the total luminosity.  The layer also cannot lie in
front of the optically visible H~{\sc ii} region, which covers most or a
considerable fraction of it, because its short wavelength emission is
well detected and the extinction over it does not show much variation.

We, therefore, favour a configuration in which the near- and
mid-IR emission come from one, or more likely, several optically
hidden sources.  Let there be $N_{\rm hid}$ of them with a
luminosity $L_{\rm hid}$ with $N_{\rm hid} L_{\rm hid} \sim 0.27
L_{\rm gal}$ for each.  For each, one has to do a radiative transfer
calculation to compute their spectra.  They can then be added
independently to the optical and FIR fluxes.  These dust hidden
source(s) may be O star clusters, which are still dust enshrouded and at an
earlier evolutionary stage than the visible super-star clusters.  Although
there must be some general trigger for the star burst in SBS~0335$-$052E,
one sees six separated clusters and their ignition is not likely to
have occured simultaneously within a period much less than the
lifetime of an O6 star.

The hidden sources of total luminosity $N_{\rm hid}L_{\rm hid}$ may be
anywhere within the field of view of the detector: within or outside
the H~{\sc ii} region, in front of or behind it. It does not make much
difference whether one imagines a few or a greater number of
(spherical) hidden sources. They are in many ways quite similar: they
are small ($< 10$ pc), so that they would never noticably obscure the
H~{\sc ii} region and have a visual optical depth $\tau_{V} \sim 6$.
The model that enters Fig.\ref{fig15} has 40 such hidden sources, with each
having a radius of 1 pc and a luminosity $L_{\rm hid} = 2.5 \times 10^7$
$L_\odot$, representing a very dense cluster of about 100 O stars.

In Fig.\ref{fig15}, the contribution from the hidden sources
emitting in the infrared is given by the thin red curve peaking 
near 25 $\mu$m.
The cold component is also shown.  To improve the fit, another warm
component with $T_2= 160$ K and a dust mass $M_2 \sim 15$ $M_\odot$ is
shown.  One would find a temperature of $\sim$160 K at a
distance of 10 pc from an unobscured source of $10^9$ $L_\odot$.

We mention that a satisfactory overall fit to the SED can, in
principle, be achieved if silicate dust is absent, so that there is
only carbon dust and no 10$\mu$m feature.  Then one does not need the
hidden sources, and the dip near 10$\mu$m would result from a crossover
of two temperature components. However, the heavy elements in this galaxy
come from exploding massive stars, and it is hard to avoid the creation of
silicon and oxygen.

The existence of an additional very cold component ($\ll 60$K) can
never be excluded, but it sets limits on its maximum mass because the
emission at 0.8 mm is clearly free-free radiation.  For example, the
mass of a 20K component, which would be about 3 kpc away from the
galaxy center, could not be much more massive than $10^4$ $M_\odot$.

%%%%%%%%%%%%%%%%%%%%%%%%%%%%%%%%%%%%%%%%%%%%%%%%%%%%%%%%%%
% Endrik Fig.15
%%%%%%%%%%%%%%%%%%%%%%%%%%%%%%%%%%%%%%%%%%%%%%%%%%%%%%%%%
%%%%%%%%%%%%%%%%%%%%%%%%%%%%%%%%%%%%%%%%%%%%%%%%%%%%%%%%%

\section{Summary \label{sum}}

We studied global characteristics of dust emission in a large sample of 
emission-line star-forming galaxies. The sample consists of two subsamples.
One subsample includes $\sim$4000 compact star-forming galaxies from
the SDSS, which were also detected in all
four bands at 3.4$\mu$m, 4.6$\mu$m, 12$\mu$m, and 22$\mu$m of the
WISE all-sky survey. The second subsample is a 
sample of 28 star-forming galaxies observed with Herschel in the FIR
range \citep{RR13}. For the second sample, we also use observations by 
WISE and Spitzer in the MIR range and various observations in 
sub-mm and radio ranges. Both SDSS and Herschel samples are supplemented 
by available data in the UV, optical, and NIR ranges.

For the SDSS sample, we use 12$\mu$m and 22$\mu$m
WISE fluxes to determine temperatures, luminosites, and masses of 
warm dust. For the Herschel sample, we adopt a
three component dust model to derive global parameters for cold, warm, and hot
components. The parameters of cold and warm dust are compared with 
other global parameters,
such as stellar masses and luminosities in the UV and optical ranges.

Our main results are as follows.

1. We derived temperatures, luminosities, and masses of the warm dust
in SDSS compact star-forming galaxies, adopting a modified black-body
distribution with a dust emissivity index $\beta$=2.0. The average
temperature of warm dust is $\sim$ 120K
and does not depend on the galaxy metallicity. 
It is slightly larger for smaller $\beta$.
Warm dust luminosities are strongly correlated with H$\beta$ luminosities,
implying that the main source of warm dust heating is 
radiation of star-forming regions that include ionising UV radiation,
and, thus, warm dust is located inside and/or near H~{\sc ii} regions.

2. Warm dust masses in SDSS compact galaxies weakly depend on the total stellar
masses, but they are much more tightly correlated with masses of young
stellar populations. We find a strong decrease of 
warm-dust-to-young-stellar-population mass ratio 
with decreasing equivalent width EW(H$\beta$) of the
H$\beta$ emission line, implying that this ratio is higher in younger
star-forming regions and that warm dust is mainly heated by ionising radiation
of H~{\sc ii} regions. We also varied $\beta$ and found that warm dust masses 
would be higher by a factor of $\sim$2 with $\beta$ = 1.0,
as compared to those with $\beta$ = 2.0.

3. We used WISE, Spitzer, and Herschel photometric data, as well as the
optical and Spitzer/{\it IRS} spectra to derive global dust characteristics 
in 28 compact galaxies
from the Herschel sample \citep{RR13} by adopting a three-component dust
model and modified blackbody emissivity with the emissivity index 
$\beta$ = 2.0. The mean dust temperatures are $\sim$ 30K and $\sim$ 90K
for the cold and warm components, respectively, with no
clear dependence on metallicity. Since the emission of hot dust may not
be in equilibrium, we do not analyse hot dust parameters because they
may deliver erroneous results.
There are two outliers, the two sources with lowest metallicity, 
SBS~0335$-$052E and I~Zw~18, which have higher cold dust temperatures.
We find general agreement between
the cold dust temperatures derived by \citet{RR13} and in this paper.

4. Similarly to the SDSS sample, we find a tight relation between the warm dust
and H$\beta$ luminosities for galaxies from the Herschel
sample. Additionally, correlations are found between cold dust
and H$\beta$ luminosities. All these relations
suggest that the main source of dust heating in star-forming galaxies is
UV radiation of young stellar populations. 
%\textbf{We also found that on average
%the fraction of (cold, warm and hot - remove?) dust emission is increased 
%with increasing H$\beta$ luminosity.}

5. The masses of the cold dust in the Herschel sample galaxies are in 
the range of $\sim$ 10$^2$ -- 10$^6$ $M_{\odot}$ and on average they
are by a factor of $\sim$10$^3$ higher than those of the warm dust.
We found a tight correlation between masses of cold and warm dust,
implying the same source of dust heating, UV radiation of star-forming regions.
It is proposed to use the relation between warm and cold dust masses for
estimations of the total dust mass in star-forming galaxies
with an accuracy better than $\sim$ 0.5 dex. This can appreciably increase the
number of star-forming galaxies with derived dust masses from several
tens to several thousands, because the data in the mid-infrared range
(e.g. WISE survey) used for the determination of the warm dust
characteristics are numerous.

6. We find that dust-to-neutral gas and dust-to-metal mass ratios 
in Herschel 
star-forming galaxies strongly decline with decreasing metallicity, 
which is similar
to results from other studies of local emission-line galaxies,
high-redshift gamma-ray burst (GRB) hosts, and damped Ly$\alpha$ absorbers
(DLAs). On the other hand, the dust-to-ionised gas mass ratios are
about one hundred times as high, indicating that most of dust is
located in the neutral gas.
%\textbf{This finding can be explained 
%by the fact that considered galaxies are unevolved systems with small
%fraction of mass in stars, all the rest of the mass is in gas.}****Is it need?

7. It is found that thermal free-free emission of ionised gas in star-forming
galaxies is important in the sub-mm range and may cause the sub-mm excess,
as discussed by \citet{RR13}. This effect is stronger in galaxies with
lower metallicity and higher star-formation rate.

8. We showed that the optically thick foreground layer invoked by 
\citet{PS02} and \citet{H13} to explain silicate dust absorption at 9.7 $\mu$m
in SBS~0335$-$052E is in conflict with observations in the UV, optical, and 
radio ranges and results in much higher radio emission than that observed.
We propose a model in which hidden star formation in
SBS~0335$-$052E resides in one or several dense and compact gas clouds,
which are optically thick at 9.7 $\mu$m. These hidden star clusters emit
only in the infrared range and do not produce an emission excess in the radio 
range, leading to satisfactory agreement with all observational data.

\acknowledgements
%We thank the referee S. Bianchi for valuable comments.
Y.I.I., N.G.G. and K.J.F. are grateful to the staff of the Max Planck 
Institute for Radioastronomy (MPIfR) for their warm hospitality. 
Y.I.I. and N.G.G. acknowledge financial support by the MPIfR.
% and the support of the Cosmomicrophysics project of the National
%Academy of Sciences of Ukraine. 
GALEX is a NASA mission managed by the Jet Propulsion Laboratory. 
This publication makes use of data products from the Two Micron All Sky 
Survey, which is a joint project of the University of Massachusetts and the 
Infrared Processing and Analysis Center/California Institute of Technology, 
funded by the National Aeronautics and Space Administration and the National 
Science Foundation. This publication
makes use of data products from the Wide-field Infrared
Survey Explorer, which is a joint project of the University of
California, Los Angeles, and the Jet Propulsion Laboratory,
California Institute of Technology, funded by the National Aeronautics
and Space Administration. Funding for the Sloan
Digital Sky Survey (SDSS) and SDSS-II has been provided by
the Alfred P. Sloan Foundation, the Participating Institutions,
the National Science Foundation, the U.S. Department of
Energy, the National Aeronautics and Space Administration,
the Japanese Monbukagakusho, and the Max Planck Society,
and the Higher Education Funding Council for England.
This research has made use of the NASA/IPAC Extragalactic Database (NED) 
which is operated by the Jet Propulsion Laboratory, California Institute 
of Technology, under contract with the National Aeronautics and Space 
Administration. 
%\clearpage

\renewcommand{\baselinestretch}{1.0}

\Online

\setcounter{figure}{5}

%%%%%%%%%%%%%%%%%%%%%%%%%%%%%%%%%%%%%%%%%%%%%%%%%%%%%%%%%%
% fits of spectra
%%%%%%%%%%%%%%%%%%%%%%%%%%%%%%%%%%%%%%%%%%%%%%%%%%%%%%%%%
\begin{figure*}
\begin{center}
\includegraphics[width=16.0cm,angle=0]{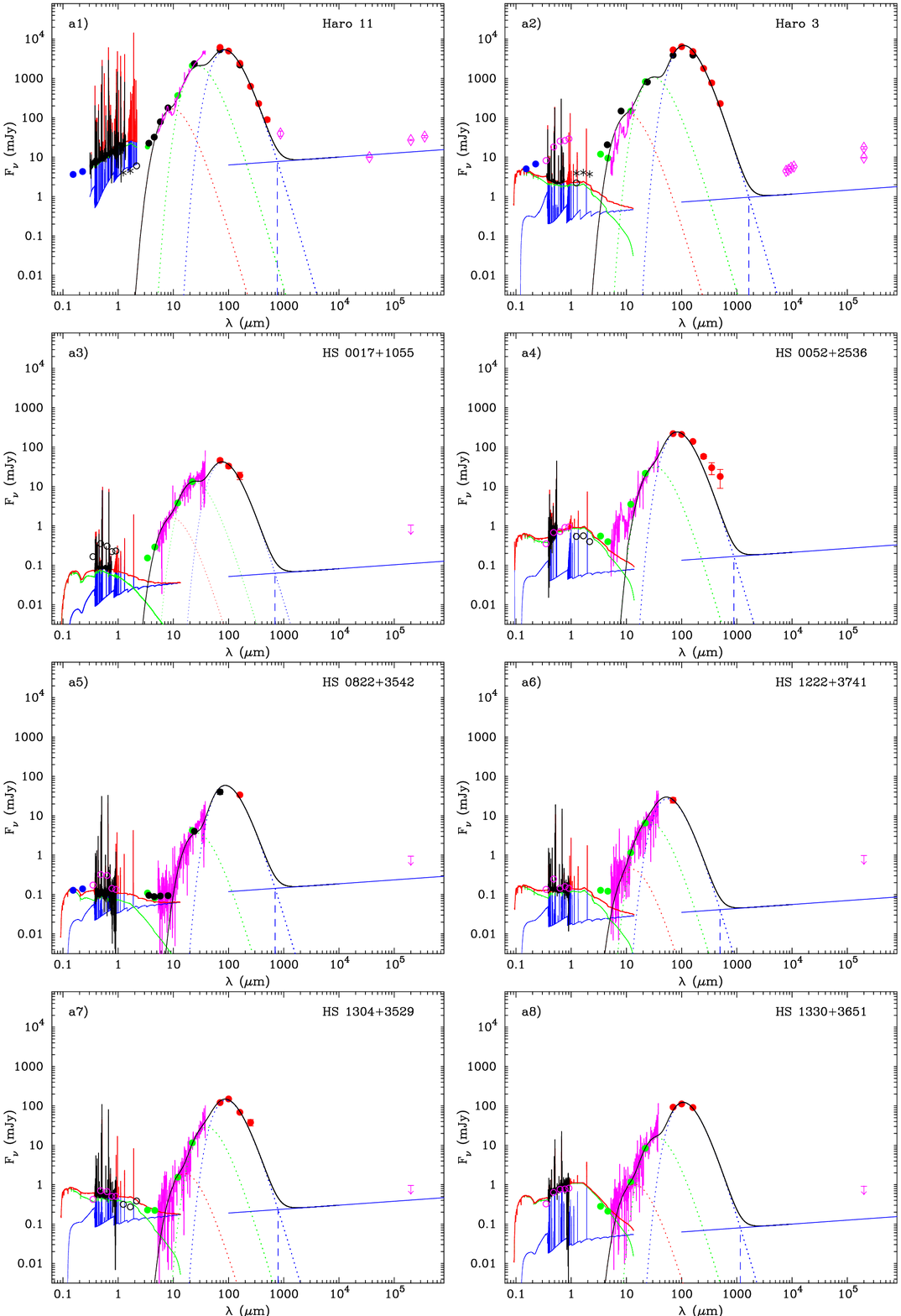}
\end{center}
\caption{\label{fig6} Spectral energy distribution fits of the
galaxies from Herschel sample. Observed spectra in the optical
and MIR ranges are shown by black and magenta lines. Photometric data
are shown from GALEX (blue filled circles),
SDSS (open magenta circles), 2MASS (black open circles), 
WISE (green symbols), Spitzer (black filled circles), Herschel 
(red filled circles), and sub-mm and radio (filled magenta circles).
Blue, green and red spectra at $\lambda$ $<$15 $\mu$m are fitted SEDs for
ionised gas, stellar and total emission, which are produced from the optical 
spectra. Fits of dust 
emission at $\lambda$ $>$ 5 $\mu$m are shown by black dotted lines for 
different components and by black solid line for total emission.
Blue solid line at $\lambda$ $>$ 100 $\mu$m is free-free emission extrapolated
from the extinction- and aperture-corrected H$\beta$ flux. Blue dashed line 
indicates the wavelength at which dust flux is equal to extrapolated 
free-free emission.}
\end{figure*}

\setcounter{figure}{5}

\begin{figure*}
\begin{center}
\includegraphics[width=16.0cm,angle=0]{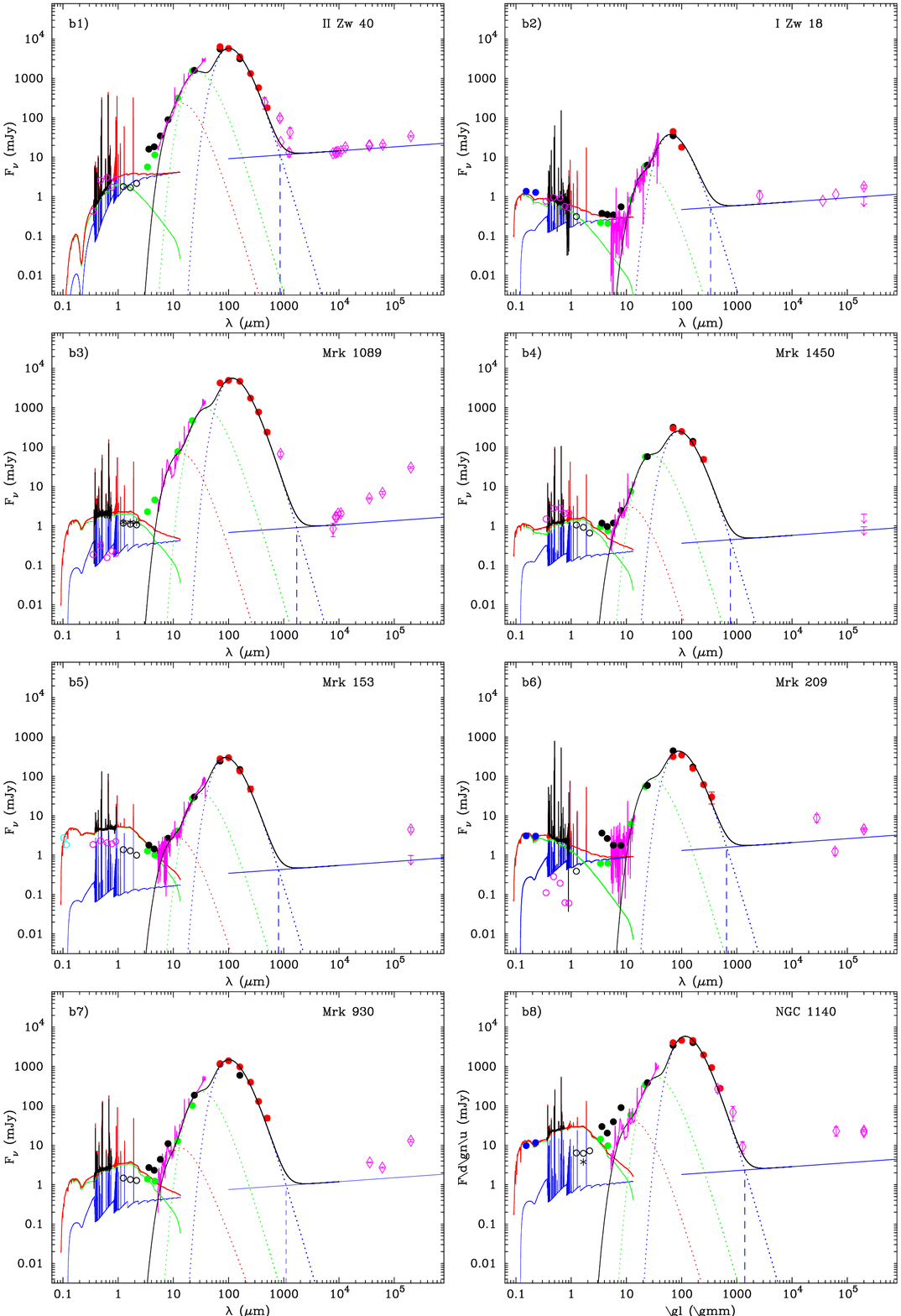}
\end{center}
\caption{Continued.  }
\end{figure*}

\setcounter{figure}{5}

\begin{figure*}
\begin{center}
\includegraphics[width=16.0cm,angle=0]{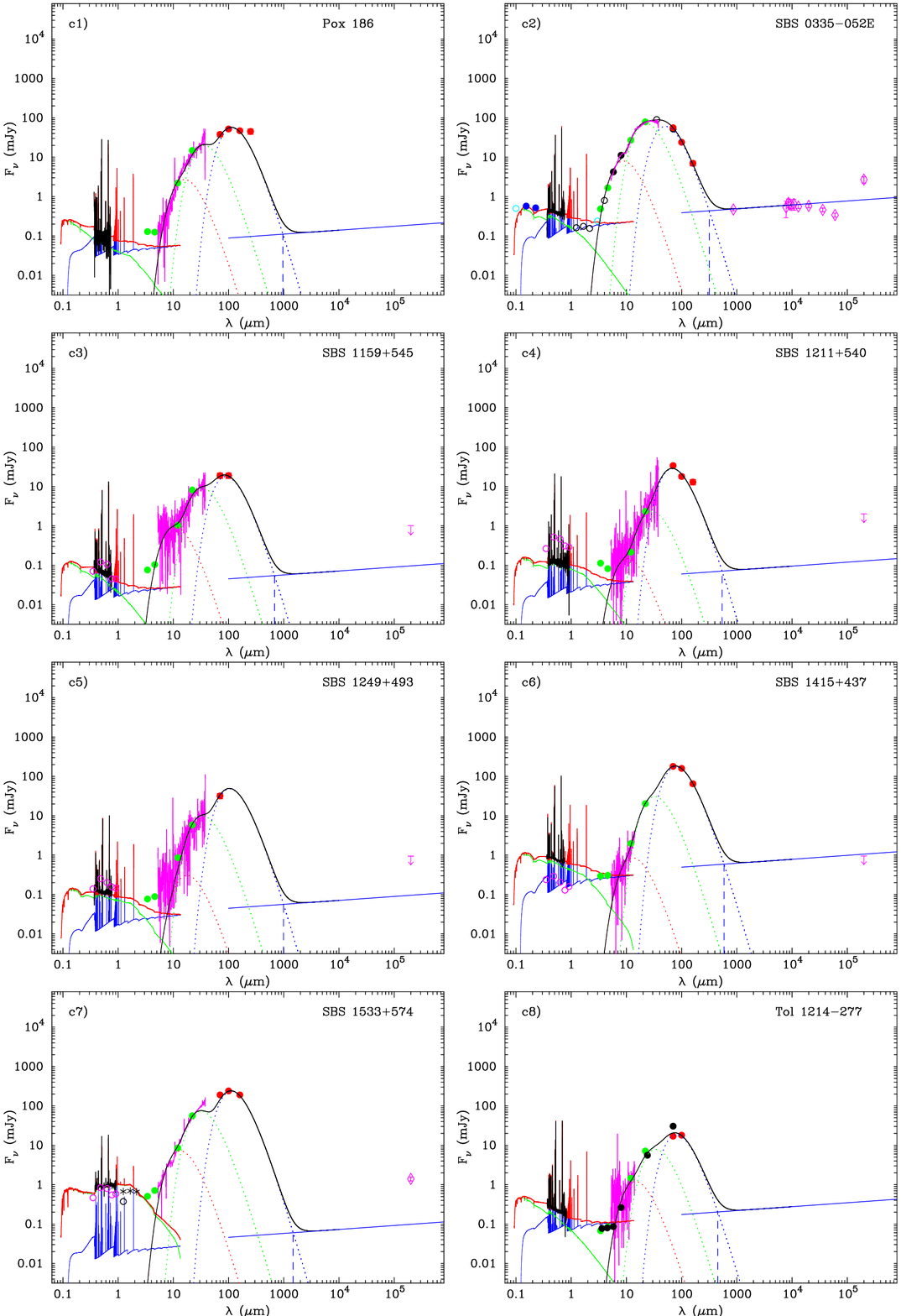}
\end{center}
\caption{Continued.  }
\end{figure*}

\setcounter{figure}{5}

\begin{figure*}
\begin{center}
\includegraphics[width=16.0cm,angle=0]{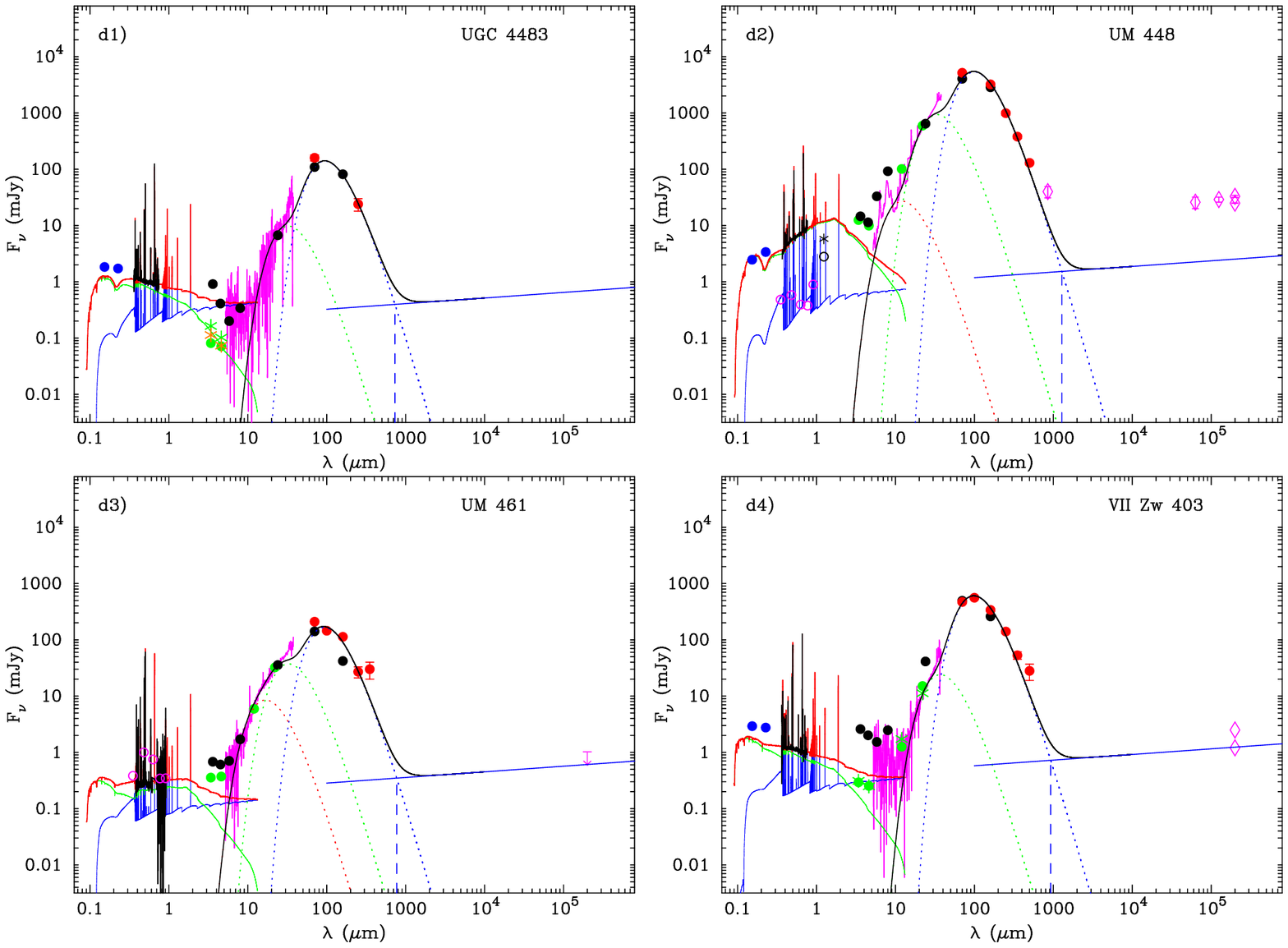}
\end{center}
\caption{Continued.  }
\end{figure*}

\end{document}